\documentclass[12pt]{elsarticle}

\usepackage{graphicx}

\usepackage{amsmath,amssymb,amsfonts, color}
\usepackage{units}
\usepackage[hang,raggedright]{subfigure}
\usepackage{setspace}

\journal{Photonics and Nanostructures}

\begin{document}

\begin{frontmatter}



\title{Design of microcavities in diamond-based photonic crystals by Fourier- and real-space analysis of cavity fields }


\author{Janine Riedrich-M\"oller}
\author{Elke Neu}
\author{Christoph Becher \corref{*}}

\address{Fachrichtung 7.3 (Technische Physik), Universit\"{a}t des Saarlandes, Campus E 2.6 \\ 66123 Saarbr\"{u}cken, Germany}

\cortext[*]{Corresponding author. Tel.: +49 (0)681 302 2466; Fax: +49 (0)681 302 4676}
\ead{christoph.becher@physik.uni-saarland.de}

\begin{abstract}
We present the design of two-dimensional photonic crystal microcavities in thin diamond membranes well suited for coupling of color centers in diamond.
By comparing simulated and ideal field distributions in Fourier and real space and by according modification of air hole positions and size, we optimize the cavity structure yielding high quality factors up to $Q=320000$ with a modal volume of $V_{\text{eff}}=0.35 (\lambda/n)^3$.  Using the very same approach we also improve previous designs of a small modal volume microcavity in silicon, gaining a factor of 3 in cavity $Q$. In view of practical realization of photonic crystals in synthetic diamond films,  it is necessary to investigate the influence of material absorption on the quality factor. We show that this influence  can be predicted by a simple model, replacing time consuming simulations.
\end{abstract}


\end{frontmatter}

\biboptions{compress}


\section{Introduction}\label{sec:Introduction}
Microcavities in two-dimensional photonic crystal slabs allow to strongly confine light in volumes smaller than one cubic wavelength.
They are expected to enable the realization of single photon emitters \cite{Englund2005}, low threshold nanolasers \cite{Strauf2006, Noda2006, Nomura2006}, ultra small filters \cite{Takano2006} and highly-efficient emitters in which  the spontaneous emission of light from single emitters is controlled at the quantum level \cite{Yoshie2004, Hennessy2007, Khitrova2006}.
Due to highly developed processing techniques for semiconductor materials most of these quantum information devices have already been demonstrated in silicon or GaAs.
Besides the use of semiconductor materials for quantum information processing, diamond has attracted significant interest in recent years due to the extraordinary properties of  optical active defect centers  that can be controlled at room-temperature \cite{Prawer2008, Greentree2008}.
There have been a number of recent proposals for employment of color centers in diamond for cavity enhanced single photon sources \cite{Su2008}, for cavity enhanced spin measurements \cite{Young2009} or  as optical qubits in quantum networks \cite{Childress2006}, quantum gates \cite{Greentree2006} and measurement-based quantum computing \cite{lim:012304, Benjamin2009}.

All these proposals require direct coupling of an emitter to a cavity mode with high quality factor $Q$ and small modal volume $V_{\text{eff}}$. In such a cavity the spontaneous emission of an emitter, placed in the maximum of the electric field of the cavity mode, is enhanced by the Purcell-Factor $F$ \cite{Purcell1946}:
\begin{equation}
F= \frac{3}{4 \pi^2}\frac{Q}{V_{\text{eff}}}\left(\frac{\lambda}{n}\right)^3
\label{Purcell-Factor}
\end{equation}
In this weak coupling regime the cavity figure of merit scales as $Q/V_{\text{eff}}$, whereas the dynamics of strong emitter-photon coupling scales as $Q/\sqrt{V_{\text{eff}}}$ \cite{Kreuzer2008}.

In recent years, several photonic crystal cavity designs have  been proposed, e.g. modified line- or point defects in  two-dimensional photonic crystal membranes.
In silicon, modified waveguides structures have  been fabricated yielding very high quality factor of $ 10^5 $ - $2.5\cdot 10 ^6$ \cite{Bong-Shik2005, Eiichi2006, Takahashi2007} with modal volumes of 1.2 - \unit{2}($\lambda/n$)$^3$. For diamond-based modified waveguide structures similar  theoretical $Q$-factors  with a modal volume of about $V_{\text{eff}} \approx 1.7(\lambda/n)^3$ have been predicted \cite{Bayn2006, Bayn2008, Tomljenovic-Hanic2009}. In practice, the cavity $Q$-factor in diamond might strongly be limited by optical loss in the diamond material \cite{Wang2007}.

In order to obtain large enhancement of the spontaneous emission rate for cavities with modest $Q$-factors, point defect cavity structures with even smaller modal volumes as compared to waveguide section cavities are required.
Different point defect structures with  zero, one or more missing holes have been proposed \cite{Zhang2004}.
As shown by Akahane et al. \cite{Yoshihiro2003} the quality factor of a point defect cavity can be improved further by optimizing the surrounding photonic crystal structure. They have demonstrated that the $Q$-factor of a photonic crystal in silicon with three missing holes (M3-cavity) can be increased up to $4.5\cdot 10^4$  \cite{Yoshihiro2003} by optimizing the next neighboring holes according to the method of ``gentle confinement''  and up to 10$^5$ by additionally shifting  the third neighboring holes outwards  \cite{Yoshihiro2005}.
The same approach has also been used to increase the $Q$-factor of a simple M1-cavity \cite{Zhang2004} in diamond with one missing hole in the center from $\approx 200$ to $3 \cdot 10^4$ \cite{S.2006} by optimizing the next neighboring holes in the $x$- and $y$-direction and up to $7 \cdot 10^4$ \cite{Kreuzer2008} by fine-tuning the holes farther outwards. Together with a modal volume of $V_{\text{eff}} =$ \unit[1.1]{$(\lambda/n)^3$}  a theoretical Purcell-Factor of $4580$ can be achieved by coupling  a color center in diamond to the optimized cavity structure.

In order to increase the Purcell-Factor even further, here we  consider a cavity design with very small modal volume: By shifting two adjacent holes outwards a so-called M0-cavity is introduced \cite{Zhang2004}. Extending the design approach of ``gentle confinement'' to remote holes and by comparing simulated and ideal field distributions in Fourier and real space, we show that the  cavity $Q$-factor is improved significantly by optimizing the surrounding holes not only in the close vicinity of the point defect but also farther away. Exemplarily, we consider in detail the design process of a M0-cavity in diamond  and in silicon.

Due to significant progress  in diamond  processing techniques, first experimental demonstrations of cavity modes in photonic crystal defect cavities in diamond have been achieved \cite{Wang2007}.  However, the measured quality factors in diamond are  more than one order of magnitude smaller than the theoretically predicted values.
This limitation might be  due to scattering losses from the grain boundaries and subsequent emission from a nonsmooth surface. Another loss mechanism is  material absorption of the nano-crystalline diamond slab. In our simulations, we investigate the influence of material absorption on the quality factor. We show that this influence can be approximated by two simple models, replacing time-consuming simulations.

One possibility to fabricate thin films with enhanced optical quality is the use of single-crystal diamond membranes. Recently,  single-crystal free-standing membranes  \cite{Olivero2005, Olivero2006, Fairchild2008} and waveguide structures \cite{Hiscocks2008, Hiscocks2008a} in  diamond have been produced using focused ion  beam milling. The use of such single-crystal diamond membranes with small absorption losses might pave the way for realization of high-$Q$ photonic crystal microcavities.

\section{Cavity design}\label{sec:CavityDesign}
Our  work is focused on two dimensional photonic crystals consisting of a triangular lattice of air holes  with a lattice constant $a$ in a thin membrane of diamond  suspended in air. The periodic structure gives rise to a photonic band gap for TE-like modes \cite{Kreuzer2008, Bayn2006}.  By introducing a defect, light can be localized in three dimensions within  small volumes: in the horizontal plane, light is localized due to distributed Bragg reflection, and in the vertical direction due to total internal reflection.

\subsection{Computation method}\label{sec:ComputationMethod}
The calculation of the near field patterns of the cavity mode was
performed with a finite difference time domain algorithm (FDTD)
\cite{Taflove1998}, using a freely available software package with
subpixel smoothing for increased accuracy \cite{Farjadpour2006}. The
simulated PC structure is a 27$a$ $\times$ 27 $a$ super cell with
height 8 $a$, surrounded by perfectly absorbing boundary conditions
(PML) \cite{Berenger1994}. To extract the resonance frequency as well
as the cavity $Q$-factor, we use a filter diagonalization method
\cite{Mandelshtam1997}. The mode volume $V_{\text{eff}}$ is
calculated by \cite{Vuckovic2001}:
\begin{equation}
V_{\text{eff}}= \frac{\int  \epsilon(\vec{r})\,
|\vec{E}(\vec{r}) |^2 d^3 r}{\max {\epsilon(\vec{r})
|\vec{E}(\vec{r})|^2}} \label{Veff}
\end{equation}
The integral of the electric energy is taken over the whole computational cell.
 To reduce the simulation time and the amount of stored data
 we implement mirror boundary conditions in the $x$, $y$ and $z$ directions. To assure that the employment of mirror boundary
conditions does not effect the calculated $Q$-factors, we reanalyze the structure at the beginning and at the end of our design process by applying an even mirror symmetry to the $z=0$
plane only in order to select TE-like modes. Calculations are initially performed with a resolution of 32 points per lattice constant. In order to check for errors due to numerical discretization we also use a resolution of 40 and 50 points per lattice constant for all optimized structures of Sec. \ref{sec:optimization} but find similar $Q$-factors.

The $Q$-factor calculated by the FDTD algorithm might vary with the chosen resolution of the simulation. Therefore, it is interesting to compare the FDTD results with an alternative method which calculates the $Q$-factor via the stored energy in the cavity and the radiated power \cite{Taflove1998, Vuckovic2001, Dirk2005,   Vuckovic2002b}. This second method permits to draw conclusions   about further improvement of the cavity geometry by analyzing the wave vector components inside the ``light cone'' (see Sec. \ref{sec:optimization}).
If the defect cavity is surrounded by a sufficiently large number of air holes, the cavity lifetime is mainly limited by radiation losses in the vertical direction \cite{Vuckovic2001}. The radiated power in the far field is then determined by the  2D Fourier transforms FT$_2$ of the  near field components $E_x$, $E_y$, $H_x$, $H_y$ in a plane $S$ at a distance $\Delta z$ above the photonic crystal slab \cite{Taflove1998, Vuckovic2002b}:
%
\begin{eqnarray}
    P = \frac{\eta}{8 \lambda^2 k^2}  \int \limits_{|\vec{k}_{\parallel}|\leq k } I dk_x
    dk_y   \\ \text{with} \quad
    I = \left|\text{FT}_2(H_y) + \eta\,
    \text{FT}_2
    (E_x)\right|^2 + \left|\text{FT}_2(H_x) - \eta\,\text{FT}_2
    (E_y) \right|^2, \nonumber
    \label{Fourier-vuck}
\end{eqnarray}
where $\eta$ is $\sqrt{\mu_0/\epsilon_0}$ and $I$ is the radiated intensity. The integral runs over all $\vec{k}$-vectors inside the ``light cone'' $|\vec{k}_{\parallel}| \leq k$.
The cavity near field  is computed using FDTD simulations after 400 time-steps. After that time transient effects, resulting from a cutoff of our excitation pulse, have died away.
The 2D Fourier transforms  of the near field pattern are taken at a plane $S$ in a distance $\Delta z = \lambda/2$ \cite{Vuckovic2002b} above the photonic crystal slab.
To compare the radiation losses of different cavity designs, we normalize the radiated intensity $I$ to the stored energy $U$ in the cavity where $U$ is given by \cite{Dirk2005}:
\begin{equation}
U=  \frac{1}{2}\int (\epsilon_0 \epsilon(\vec{r}) |\vec{E}(\vec{r})|^2 + \mu_0|\vec{H}(\vec{r})  |^2 )d^3r
\label{storedEnergy}
\end{equation}
The quality factor $Q$ is  defined by the stored energy $U$ in the
cavity divided by the total radiated Power $P$ per cycle \cite{Dirk2005}:
\begin{equation}
    Q= \omega \frac{U }{ P},
    \label{Q_U/P}
\end{equation}
where $\omega$ denotes the angular frequency of the cavity mode.
To  calculate the $Q$-factor, the integral \eqref{storedEnergy} of the  energy stored between two planes $S$ above and below the photonic slab  as well as the integral in equation \eqref{Fourier-vuck} are performed as a discrete sum. Finally we take the time average of the radiated power $P$ and the stored energy $U$ over one period. This final time-averaging can be avoided by calculating the imaginary part as well as the real part of the fields instead of the real part only. It is sufficient to evaluate the complex fields at one instant in time, since the energy of complex fields does not change over one period.

\subsection{Starting point: simple M0-cavity}
\label{sec:M0Cavity}

\begin{figure}
  \centering
  \subfigure[M0-cavity design]{
   \includegraphics[height=0.197\textheight]{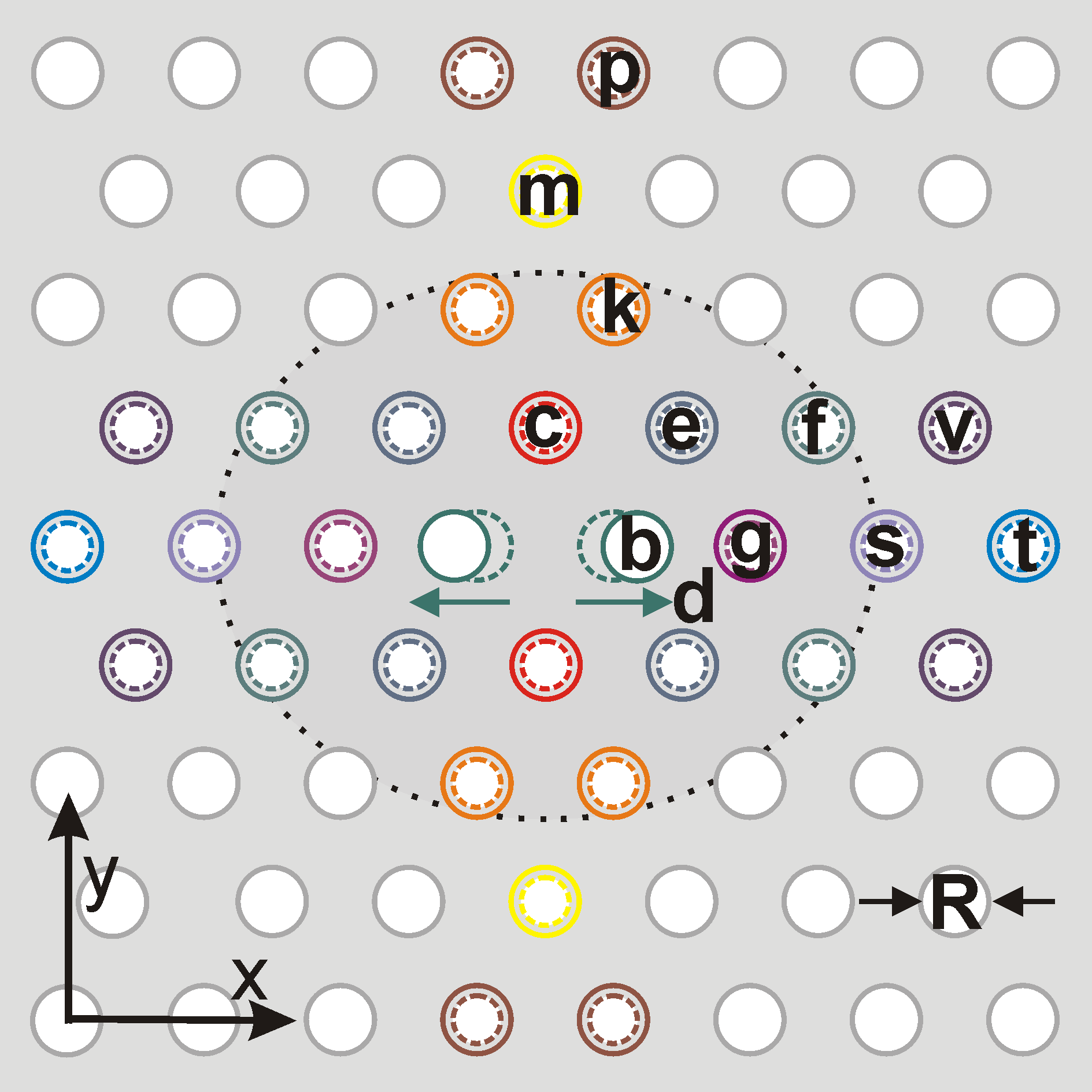}
  \label{fig:M0-design}}
  \subfigure[$E_y$-field]{
   \includegraphics[height=0.197\textheight]{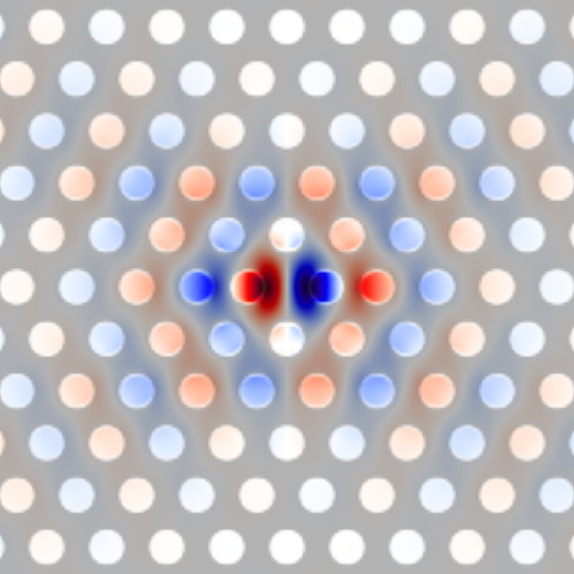}
  \label{fig:M0Ey}}
    \subfigure[$H_z$-field]{
   \includegraphics[height=0.197\textheight]{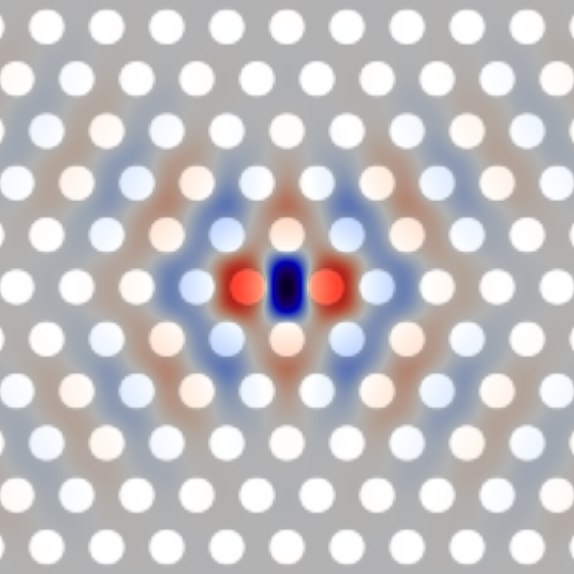}
  \label{fig:M0Hz}}
  \caption{Design and field components of the M0-cavity: (a) Nomenclature of the holes surrounding the defect. (b) $E_y$- and (c) $H_z$-component of the M0$^A$-cavity mode.
  \label{fig:M0-design+Felder}}
\end{figure}

As a basic cavity design, we choose a point defect structure with a very small modal volume: By shifting two adjacent holes along the $x$-direction by a distance $d$, a so called M0-cavity is created \cite{Zhang2004}.
Figure \ref{fig:M0-design} shows the nomenclature of the M0-cavity design, which will be used in the following. As a first example, we consider a M0-cavity in a lossless diamond membrane with a refractive index  of $n=2.4$ \cite{Zaitsev2001}. The cavity design supports only one resonant mode, whose   $E_y$ and  $H_z$-components are shown in Fig. \ref{fig:M0Ey}-\ref{fig:M0Hz}, respectively.
The quality factor of this simple M0-cavity depends on three
parameters: the background radius $R$ of air holes, the slab thickness $h$, and
the shift $d$ of the two holes creating the defect. By successively modifying these parameters in steps of $0.01 a$ within ranges $R \in [0.25a , \, 0.29a]$, $h \in [0.9a, \, 0.95a]$ and $d \in [0.13a, \, 0.17a]$, we find Parameter Set M0$^A$, listed in Tab. \ref{tab:M0a}, yielding the highest quality factor of $Q \approx 24000$ of all simple M0-cavities with a modal volume $V_{\text{eff}}=0.390 (\lambda/n)^3$ at a resonance frequency $\omega = 0.3652 \,(2\pi c/a)$. The quality factor seems to be independent of the slab thickness $h$ in this parameter range. Therefore, we keep $h=0.91a$ for the remainder of the paper.
\begin{table}[h]
\centering
\begin{tabular}{|c|c|c|c|c|c|}
\hline
{$R$}($a$) & {$h$}($a$) &  {$d$}($a$) &    {$Q$} & {$\omega(2\pi c/a)$} & {$ V_{\text{eff}}(\lambda/n)^3$ } \\
\hline
      0.26 &  0.91&     0.15 &      24000 &     0.3652 &       0.390 \\
\hline
\end{tabular}
\caption{Starting point for the design optimization of the
M0-cavity (Parameter Set M0$^{A}$)\label{tab:M0a}}
\end{table}


\subsection{Optimization of quality factor}
\label{sec:optimization}

The $Q$-factor of the simple M0$^{A}$ cavity is more than a factor of four smaller than equivalent designs in silicon membranes \cite{Zhang2004} due to the lower refractive index of diamond.
In order to achieve comparable $Q$-factors in diamond, the cavity design has to be optimized, without delocalizing the cavity mode, i.e. without increasing the modal volume.
In two dimensional photonic crystals light can be localized in three dimensions due to Bragg reflection in the horizontal plane and by total internal reflection (TIR) in the vertical direction by the air cladding. However, only plane wave components with in-plane wave vectors $|k_{\parallel}|> k_0$ are guided within the slab by TIR whereas all modes with wave vectors inside the ``light cone'' $|k_{\parallel}|\leq k_0$ are radiated into the air cladding according to Snell's law \cite{Yoshihiro2003} . These vertical radiation losses are crucial for photonic crystals in diamond because of the relatively low refractive index. The lower refractive index results in a higher mid-gap frequency of the band gap compared to silicon-based photonic crystals and therefore results in a larger  corresponding light cone \cite{Bayn2006}.
A suitable approach to minimize these radiation losses in the vertical direction  is the method of ``gentle confinement'' \cite{Yoshihiro2003, Yoshihiro2005}: Fourier- and real-space analysis reveals that radiation losses can be reduced by tailoring the cavity mode profile to resume a Gaussian envelope function. In the case of photonic crystals, such ``gentle confinement'' can be obtained by adjusting the neighboring holes around the introduced defect, e.g. by reducing the radii or shifting some holes outwards.
The design process of the M0-cavity  is structured as follows:

\begin{description}
  \item [Step 1] Optimization for background radius $R=0.26a$
  \begin{enumerate}
  \item [(a)] As a first optimization step, we fine-tune the holes $c$, $e$, $k$ and $b$, $g$, $f$ in the vicinity of the defect highlighted by the shaded region in Fig. \ref{fig:M0-design}, for a background radius $R=0.26a$, such that the cavity mode profile along the $y$- and $x$-axis, respectively, fits a Gaussian envelope.
  \item [(b)] Thereafter, we adjust the holes $m$, $p$ and $s$, $v$, $t$ farther out along the $y$- and $x$-axis, respectively.
  \end{enumerate}
  \item [Step 2] Optimization for background radii $R \in [0.27a, \, 0.29a]$
  \begin{enumerate}
  \item[(a)]As a next step, we check whether the background radius $R$ and thus the in-plane localization of the modified mode is still optimal. Therefore, we repeat step 1(a) for changed background radii $R \in [0.27a, \, 0.29a]$.
  \item [(b)] Finally, we repeat step 1(b) for a background radius $R=0.28a$.
  \end{enumerate}
\end{description}

\subsection*{Step 1.a: Optimization of the next-neighboring holes for background radius $R=0.26a$}
\begin{figure}
  \centering
  \subfigure[M0$^{A}$ cavity mode]{
  \includegraphics[height=0.197\textheight]{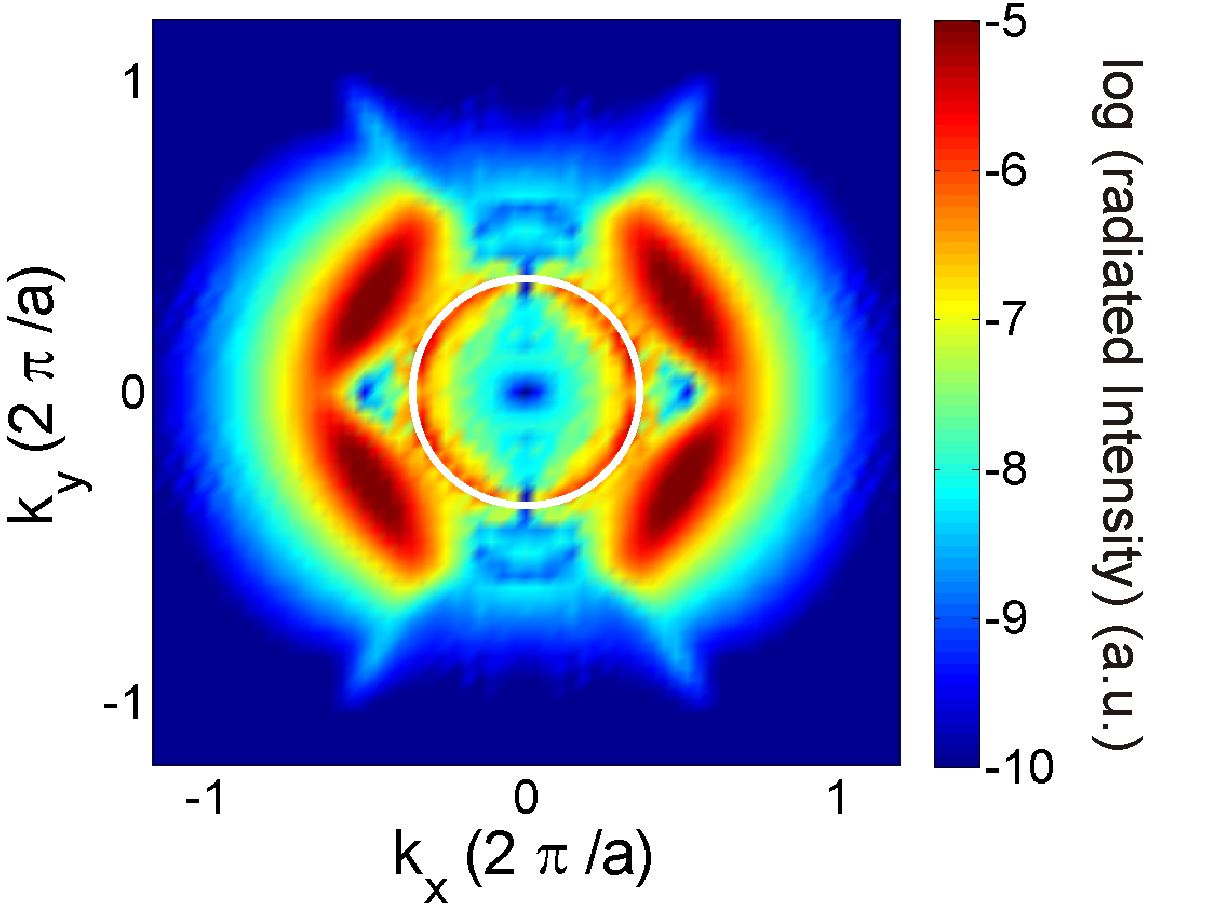}
  \label{fig:Hz-d}}
  \subfigure[M0$^{B}$ cavity mode]{
  \includegraphics[height=0.197\textheight]{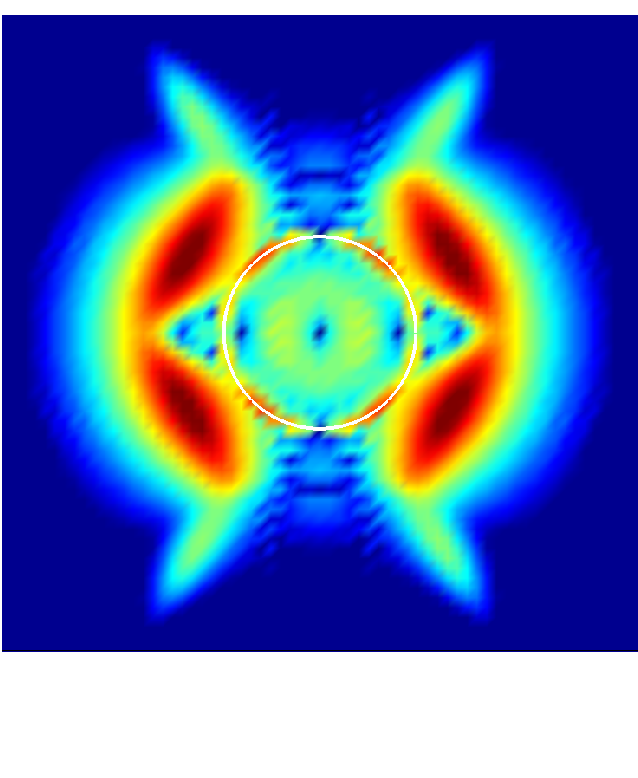}
  \label{fig:Hz-d15-Rc-Re-Rk}}
  \subfigure[M0$^{C}$ cavity mode]{
  \includegraphics[height=0.197\textheight]{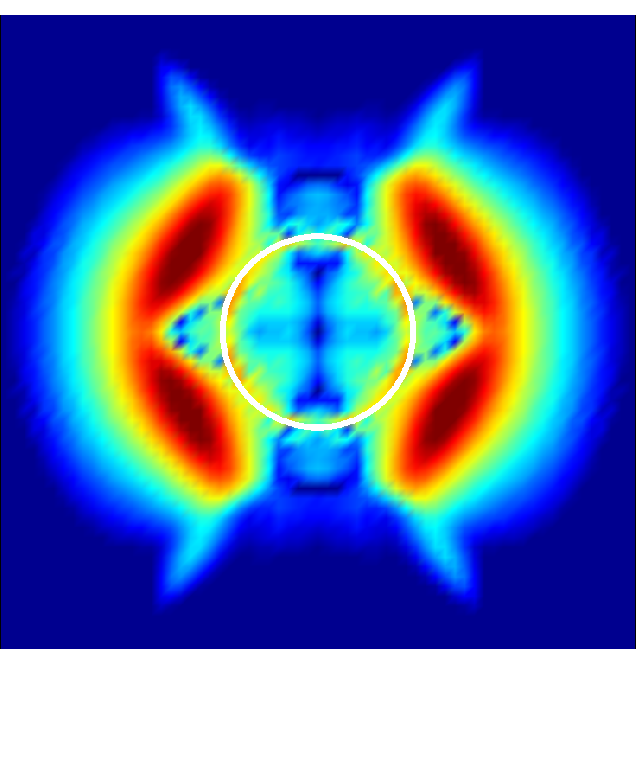}
  \label{fig:Hz-d16-Rc-Re-Rk}}
  \caption{Comparison between the radiated intensity $I$ normalized to the stored energy $U$ of (a) a simple M0$^{{A}}$ mode (b) a M0$^B$ mode, where the holes along the $y$-axis are optimized and (c) a M0$^C$ cavity mode where additionally the holes along the $x$-axis are fine-tuned.\label{fig:Vergleich-FFTHz-R,26}}
\end{figure}

The modest quality factor $Q$ of a simple M0$^A$ cavity  is explained by considering the radiation losses in the vertical direction. The radiation intensity $I$ of the M0$^{A}$ cavity calculated according to equation \eqref{Fourier-vuck} is shown in Fig. \ref{fig:Hz-d}. In order to properly compare radiation losses of different cavity designs, we normalize all plots of radiation intensity $I $ to the energy $U$ stored in the cavity.
The white circle denotes the light cone boundary $|k_{\parallel}|\leq k_0$.  Note the logarithmic color scale to visualize weak field components.
The red area inside the light cone in Fig. \ref{fig:Hz-d} indicates large leaky components: The radiated power inside the light cone divided by the stored energy is $P/U \approx 7.9 \cdot 10^{-5} (c/a)$.
We start our design process  by optimizing the cavity structure along   the $y$-axis.  In order to identify which holes have to be adjusted, we analyze the field distribution along the $y$-axis and compare it to an ideal Gaussian envelope function. Figure \ref{fig:R,26-d,15-Ex-y-Achse} shows the  $E_x$-distribution of the M0$^A$ cavity mode (black rectangles). The ideal mode profile, shown in red, is given by a sinusoidal fundamental wave multiplied by a Gaussian envelope
function (green curve in Fig. \ref{fig:R,26-d,15-Ex-y-Achse}) \cite{Yoshihiro2003}. The shaded regions indicate deviations of the cavity mode from the ideal profile. These deviations  already start in the immediate vicinity of the cavity center, thus indicating the need for optimization of the holes  $c$, $e$, $k$ (see Fig. \ref{fig:M0-design}).

As a first step, we vary the radii $R_c$, $R_e$ and $R_k$ of the holes $c$, $e$ and $k$ such that the cavity mode in the vicinity of the $y$-axis  fits a Gaussian envelope.
The optimal choice $R_c=0.23a$, $R_e=0.24a$, $R_k = 0.25a$ yields  a quality factor $Q = 71400$ with a modal volume of $V_{\text{eff}}=0.390 (\lambda/n)^3$ (Parameter Set M0$^B$ in Tab. \ref{tab:M0b}). The 2D Fourier transform of the M0$^B$-cavity is shown in Fig. \ref{fig:Hz-d15-Rc-Re-Rk}: By adjusting the holes in the immediate vicinity of the defect  the leaky components inside the light cone have been  reduced to $P/U \approx 2.8\cdot 10^{-5}(c/a)$.

\begin{table}[h]
\centering
\begin{tabular}{|c|c|c|c|c|c|c|c|}
\hline
{$R$}($a$) & {$d$}($a$)&  {$R_c$}($a$) &{$R_e$}($a$)& {$R_k$}($a$) &{$ Q$} & { $\omega(2\pi c/a)$} & {$ V_{\text{eff}}(\lambda/n)^3$ } \\
\hline
      0.26 &       0.15 &  0.23 & 0.24 & 0.25   &   71400 &     0.3622 &       0.390 \\
\hline
\end{tabular}
\caption{Optimization of first, second and third next-neighbor holes along the $y$-axis (Parameter Set M0$^{B}$).\label{tab:M0b}}
\end{table}

As a next step, we repeat the procedure above for the $x$-direction by analyzing the $E_y$-distribution along the $x$-axis as shown in Fig. \ref{fig:R,26-d,15-RcReRk-Ey-x-Achse}: The deviations from the ideal field distribution are important at the hole positions $b$, $s$ and $t$, whereas the deviations at position $g$ are quite small. The mismatch at the position of the next-neighbor holes $b$ indicates, that the displacement $d$ of the holes along the $x$-axis is no longer optimal,  after we have optimized the holes in the vicinity of the $y$-direction.
\begin{figure}
  \centering
  \subfigure[M0$^A$-cavity mode]{
  \includegraphics[width=0.43\textwidth]{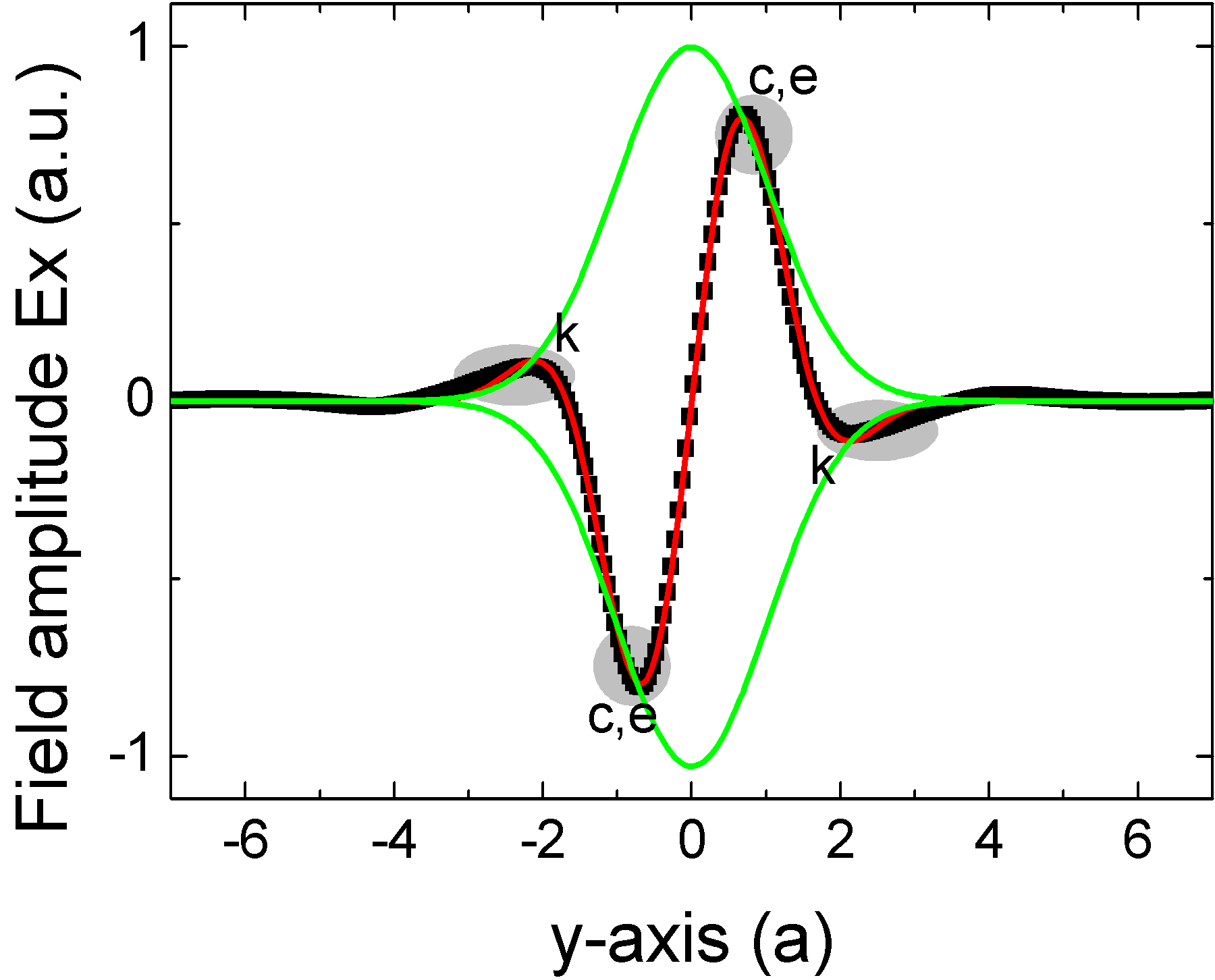}
  \label{fig:R,26-d,15-Ex-y-Achse}}
  \hspace{0.1cm}
  \subfigure[M0$^B$-cavity mode]{
  \includegraphics[width=0.43\textwidth]{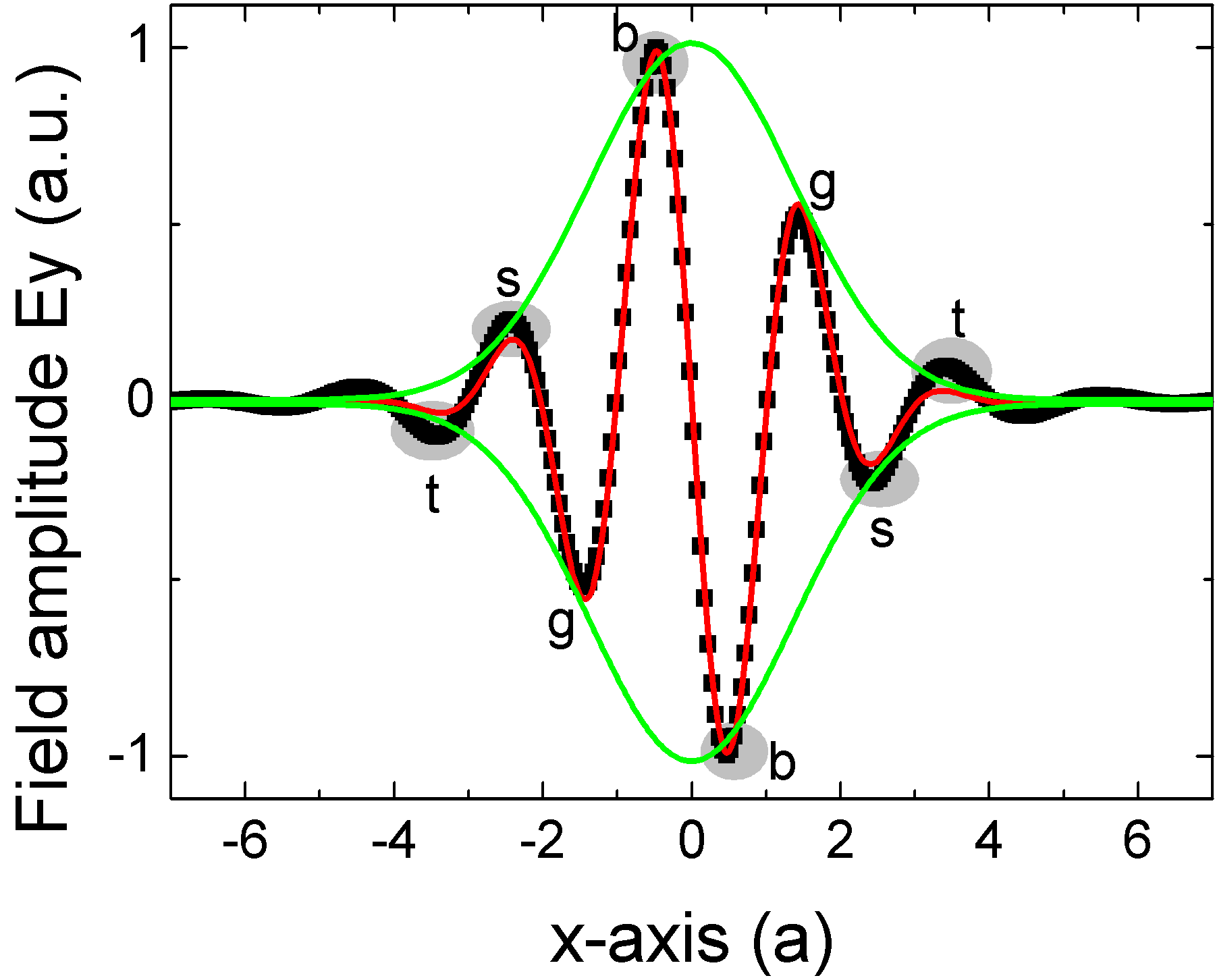}
  \label{fig:R,26-d,15-RcReRk-Ey-x-Achse}}
  \caption{Field amplitude of the M0$^A$- and M0$^B$-cavity mode along the $y$ and $x$-axis, respectively: (a) $E_x$-field of the M0$^A$-cavity along the $y$-axis: The gray regions at the hole positions $c$, $e$, $k$ indicate deviations of the calculated cavity mode (black) from the ideal mode profile (red) with a Gaussian envelope function (green). (b) $E_y$-distribution of the M0$^B$-cavity along the $x$-axis: The deviations from the ideal profile are important at positions $b$, $s$ and $t$, whereas the deviations at position $g$ are small. \label{fig:Ey-Feld-entlang-x-Achse}}
\end{figure}

By slightly increasing the shift $d=0.16a$, keeping all other parameters of the M0$^B$ structure fixed,  the quality factor can be improved significantly up to $Q=121500$ with a modal volume $V_{\text{eff}}=0.388 (\lambda/n)^3 $.
An additional modification of the radius $R_b$ of the next-neighbors does not lead to a further improvement of the cavity lifetime.
As indicated in Fig. \ref{fig:R,26-d,15-RcReRk-Ey-x-Achse} the field distribution at position $g$  seems to be already optimal. Indeed neither changing the radius $R_g$ nor shifting the holes $g$ along the $x$-axis leads to an increase of the quality factor.
Almost the same holds for the holes $f$ farther away from the defect.
By  reducing the radius $R_f$ the quality factor can be improved up to $Q=132000$.
The  cavity design  optimized in the $x$- and $y$-direction for background radius $R=0.26a$ is summarized in Parameter Set M0$^C$ in Tab. \ref{tab:M0c}. The radiation intensity $I$ is shown in Fig. \ref{fig:Hz-d16-Rc-Re-Rk}: The increase of the $Q$-factor compared to the M0$^B$ structure is attributed to a  reduction of the radiated power in the vertical direction to $P/U \approx 1.6 \cdot 10^{-5}(c/a)$.

\begin{table}[h]
\centering
\begin{tabular}{|c|c|c|c|c|c|c|c|c|}
\hline
{$R$}[$a$] & {$d$}[$a$]  & {$R_c$}[$a$] &{
$R_e$}[$a$]&  {$ R_k$}[$a$] &{$R_f$}[$a$]& {$Q$} & { $\omega[2\pi c/a]$} & {$ V_{\text{eff}}[(\lambda/n)^3]$ } \\
\hline
      0.26 &       0.16 &  0.23 & 0.24  &0.25&0.25   &   132000 &     0.3606 &       0.388 \\
\hline
\end{tabular}
\caption{Optimization of the holes along the $x$- and $y$-axis for background radius $R=0.26a$ (Parameter Set M0$^{C}$). The radii  not listed here correspond to the background radius $R$. \label{tab:M0c}}
\end{table}

\subsection*{Step 1.b: Optimization of remote holes for $R=0.26a$}
After the fine-tuning of the holes in the vicinity of the defect (see shaded region in Fig. \ref{fig:M0-design}), we consider the holes $m$, $p$ and $s$, $v$, $t$ at larger distances along the $y$- and $x$-axis, respectively.
Judging from the radiation intensity $I$ of the M0$^C$ cavity (Fig. \ref{fig:Hz-d16-Rc-Re-Rk}), the wave vector components along the $k_y$-axis are already very small. They represent  about $10\%$ of the components in the light cone. Indeed, neither changing the radii $R_m$ nor $R_p$ of the holes $m$, $p$ along the $y$-axis yields  further improvement of the cavity $Q$.

As a second step, we consider the holes $s$, $v$, $t$ along the $x$-axis. In Fig. \ref{fig:R,26-d,15-RcReRk-Ey-x-Achse}
deviations from the Gaussian envelope are visible at positions $s$ and $t$. By slightly reducing the radii $R_s=0.25a $  and $R_t = 0.25a$ the quality factor can be improved
up to $Q = 146500$ with a modal volume  $V_{\text{eff}}=0.390(\lambda/n)^3$.
Additional modification of the holes $v$ does not lead to further improvement of the cavity $Q$.
The optimal choice of the hole radii $R_m$, $R_p$ and $R_s$, $R_v$, $R_t$ in the $y$- and $x$-direction, respectively, are summarized in Parameter Set M0$^D$ in Tab. \ref{tab:M0d}. The other parameters not listed in Tab. \ref{tab:M0d} correspond to the  parameters of the M0$^C$-cavity.

\begin{table}[h]
\centering
\begin{tabular}{|c|c|c|c|c|c|c|c|}
\hline
 {$ R_m$}[$a$] &{$R_p$}[$a$] &{$R_s$}[$a$] &  {$R_v$}[$a$] &{$R_t$}[$a$]&  {$Q$} & { $\omega[2\pi c/a]$} & {$ V_{\text{eff}}[(\lambda/n)^3]$ } \\
\hline
       0.26 & 0.26  &0.25 &0.26 & 0.25  &   146500 &     0.3604 &       0.390 \\
\hline
\end{tabular}
\caption{Optimization of the holes at larger distances from the defect for background radius $R=0.26a$ (Parameter Set M0$^{D}$). The other parameters are the same as in Set M0$^C$. \label{tab:M0d}}
\end{table}

\subsection*{Step 2.a: Optimization of the next-neighbors for different background radii $R$}

At the beginning of our design process we adjusted the background radius $R$ for a simple M0-cavity. We now want to verify whether the choice of $R=0.26a$ is still optimal.
On the one hand, as discussed in Ref. \cite{Kreuzer2008}, larger background radii $R$ lead to an increase of the width of the band gap and hence to a better in-plane mode confinement. On the other hand, an increase of the radius $R$ also leads to larger wave vector components inside the light cone and thus to larger radiation losses in the vertical direction. Therefore, the choice of $R$ is a tradeoff between these two opposed contributions. By optimizing the holes around the defect,  vertical radiation losses have been   reduced significantly. We now want to check whether the in-plane confinement of the mode can be improved as well, by increasing the background radius $R \in [0.27 a, \, 0.29 a]$ in steps of 0.01$a$. Using the very same design process as described in step 1.a, we fine-tuned the holes $c$, $e$, $k$ and $b$, $g$, $f$ in the $y$- and $x$-direction, respectively, for every background radius $R \in [0.27 a, \, 0.29 a]$. The highest quality factor of an optimized cavity geometry is obtained for background radius $R = 0.28a$ (Parameter Set M0$^E$ in Tab. \ref{tab:M0e}).
For an optimized cavity structure, the confinement of the mode in the in-plane direction is considerably improved for a background radius $R=0.28a$ leading to larger quality factor $Q=226600$ and a smaller mode volume  $V_{\text{eff}}=0.350 (\lambda/n)^3$.
\begin{table}[h]
\centering
\begin{tabular}{|c|c|c|c|c|c|c|c|c|}
\hline
{$R$}[$a$] & {$d$}[$a$]  & {$R_c$}[$a$] &{
$R_e$}[$a$]& {$ R_k$}[$a$] &{$R_f$}[$a$]& {$Q$} & { $\omega[2\pi c/a]$} & {$ V_{\text{eff}}[(\lambda/n)^3]$ } \\

\hline
      0.28 &       0.16 & 0.22 & 0.24  &0.26&0.26   &   226600 &     0.3673 &       0.350 \\

\hline
\end{tabular}
\caption{Optimized M0-cavity for background radii  $R=0.28a$ (Parameter Set M0$^E$). \label{tab:M0e}}
\end{table}

\subsection*{Step 2.b: Optimization of remote holes for background radius $R=0.28a$}
\begin{figure}
  \centering
 \subfigure[M0$^{E}$-cavity mode]{
  \includegraphics[height=0.197\textheight]{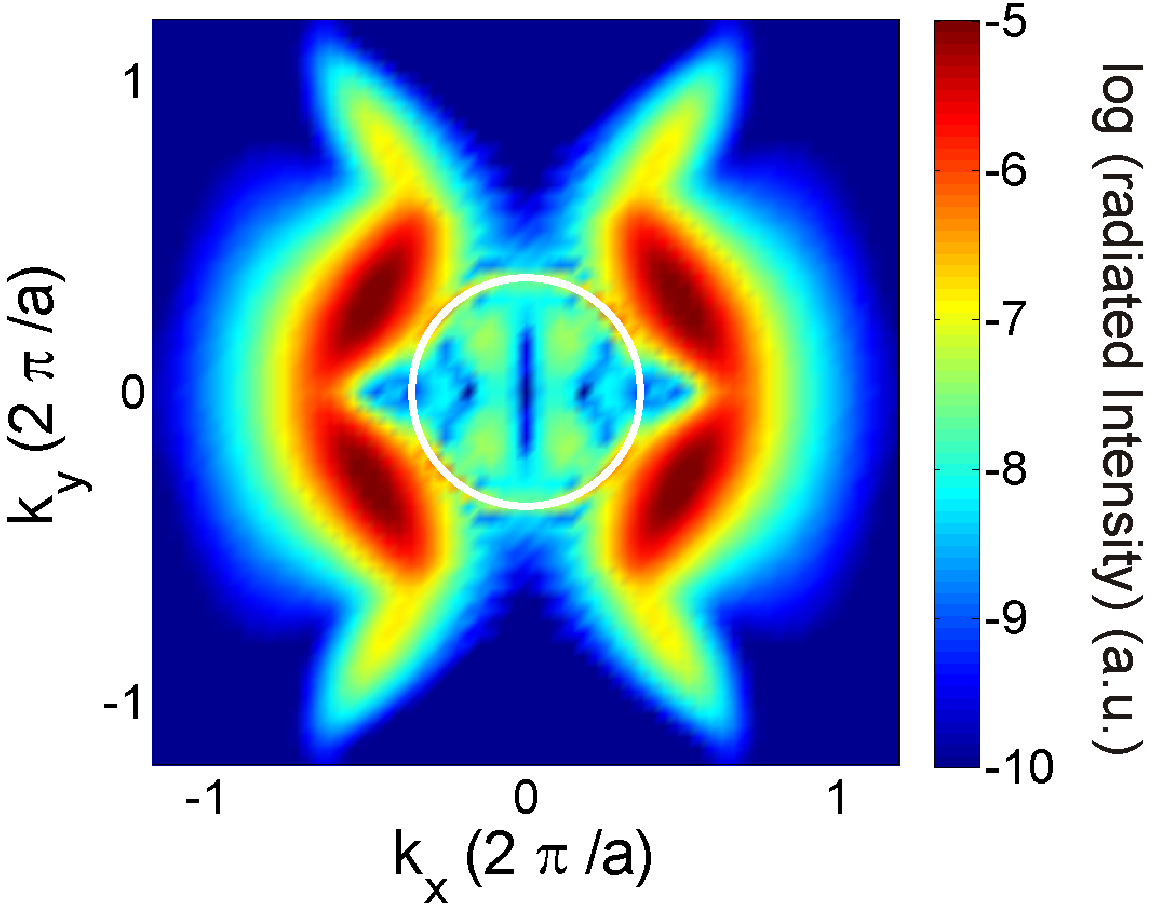}
  \label{fig:FFTHz-R28-RcReRkRf}}
  \subfigure[M0$^{F}$-cavity mode]{
  \includegraphics[height=0.197\textheight]{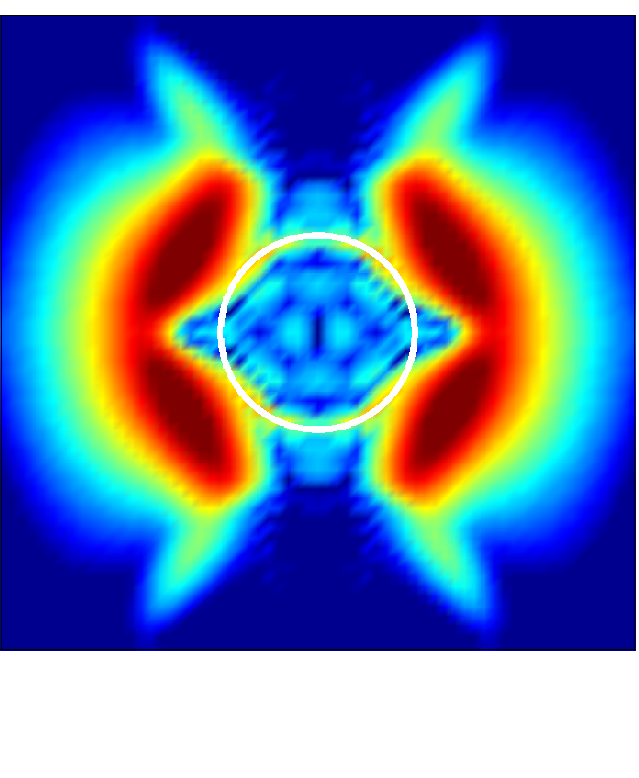}
  \label{fig:FFTHz-R28-RcReRkRfRmRpRnRsRv}}
  \hspace{0.1cm}
  \subfigure[M0$^{F}$-cavity mode]{
  \includegraphics[height=0.197\textheight]{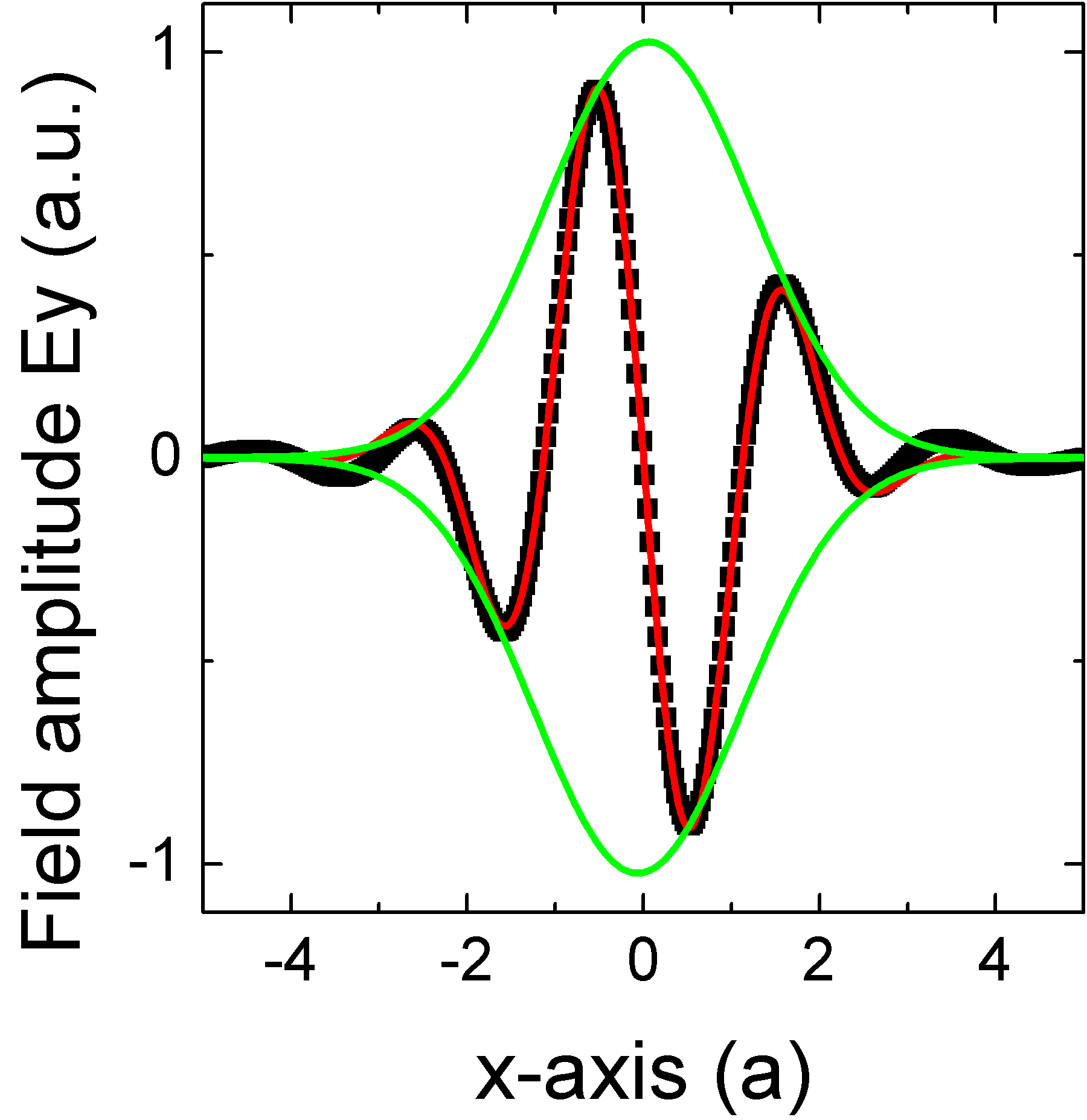}
  \label{fig:Hz-R28-RcReRkRfRmRpRsRv}}
  \caption{Comparison between the radiated intensity normalized to the stored energy for background radius $R=0.28a$ of (a) a M0$^{{E}}$ mode (b) a M0$^F$ mode, where the holes both along the $x$- and $y$-axis are optimized.  (c) $E_y$-field of the  M0$^F$-cavity mode along $x$-axis: The mode profile fits well to a Gaussian envelope. \label{fig:Vergleich-FFTHz-R,26}}
\end{figure}

After the optimization of the holes in the vicinity of the defect (see shaded region in Fig. \ref{fig:M0-design}), we again investigate the influence of the remote holes $m$, $p$ and $s$, $v$, $t$.
Fig. \ref{fig:FFTHz-R28-RcReRkRf} shows the normalized radiated intensity  of the M0$^E$ cavity mode: Wave vector components are still left inside the light cone. The radiated Power $P$  normalized to the stored energy is $P/U \approx 1 \cdot 10^{-5} (c/a)$. This suggests, that further improvement of the high-$Q$ M0$^E$-cavity can be obtained by adjusting the holes at larger distances from the defect.

Like before, we  start to optimize the holes $m$, $p$ along the $y$-axis, which leads to only slight  improvement of the $Q$-factor to 228500 for radii $R_m = 0.31a$, $R_p = 0.34a$.
As a final step, we consider the holes $s$, $v$, $t$ along the $x$-axis (see Fig. \ref{fig:M0-design}). The optimal choice of the radii $R_s=0.29a$, $R_v=0.27a$, $R_t = 0.28a$ yields another significant decrease of the leaky components inside the light cone, shown  in Fig. \ref{fig:FFTHz-R28-RcReRkRfRmRpRnRsRv}.  The reduction of radiation losses  to $P/U \approx 6.9 \cdot 10^{-6}(c/a)$ leads to a considerable increase of the quality factor up to $Q=320000$ with a modal volume $V_{\text{eff}}=0.350 (\lambda/n)^3$ (see Parameter Set M0$^F$ in Tab. \ref{tab:M0g}). Coupling a color center to such an optimized cavity leads to a theoretical Purcell-Factor of $F= 69500$.  The $E_y$-component of the M0$^F$-cavity along the $x$-axis is shown in Fig. \ref{fig:Hz-R28-RcReRkRfRmRpRsRv}: After carefully fine-tuning the holes around the defect the mode profile of the optimized structure fits well to a Gaussian envelope.

\begin{table}[h]
\centering
\begin{tabular}{|c|c|c|c|c|c|c|c|}
    \hline
        {$R_m$}[$a$]&    {$R_p$}[$a$]&  {$R_s$}[$a$]&  { $R_v$}[$a$]& {$R_t$}[$a$] &  {$Q$} & {$\omega[2\pi c/a]$} & { $ V_{\text{eff}}[(\lambda/n)^3]$ } \\
    \hline
             0.31 & 0.34 & 0.29 & 0.27& 0.28 &  320000&     0.3672 &       0.350 \\
    \hline
\end{tabular}
\caption{Optimization of the holes along the $x$-axis (Parameter Set  M0$^{F}$). The other parameters are the same as in Set M0$^E$.
\label{tab:M0g}}
\end{table}

\subsection{Fabrication tolerance tests} \label{sec:stability}
\begin{figure}
  \centering
  \subfigure{
  \includegraphics[width=0.45\textwidth]{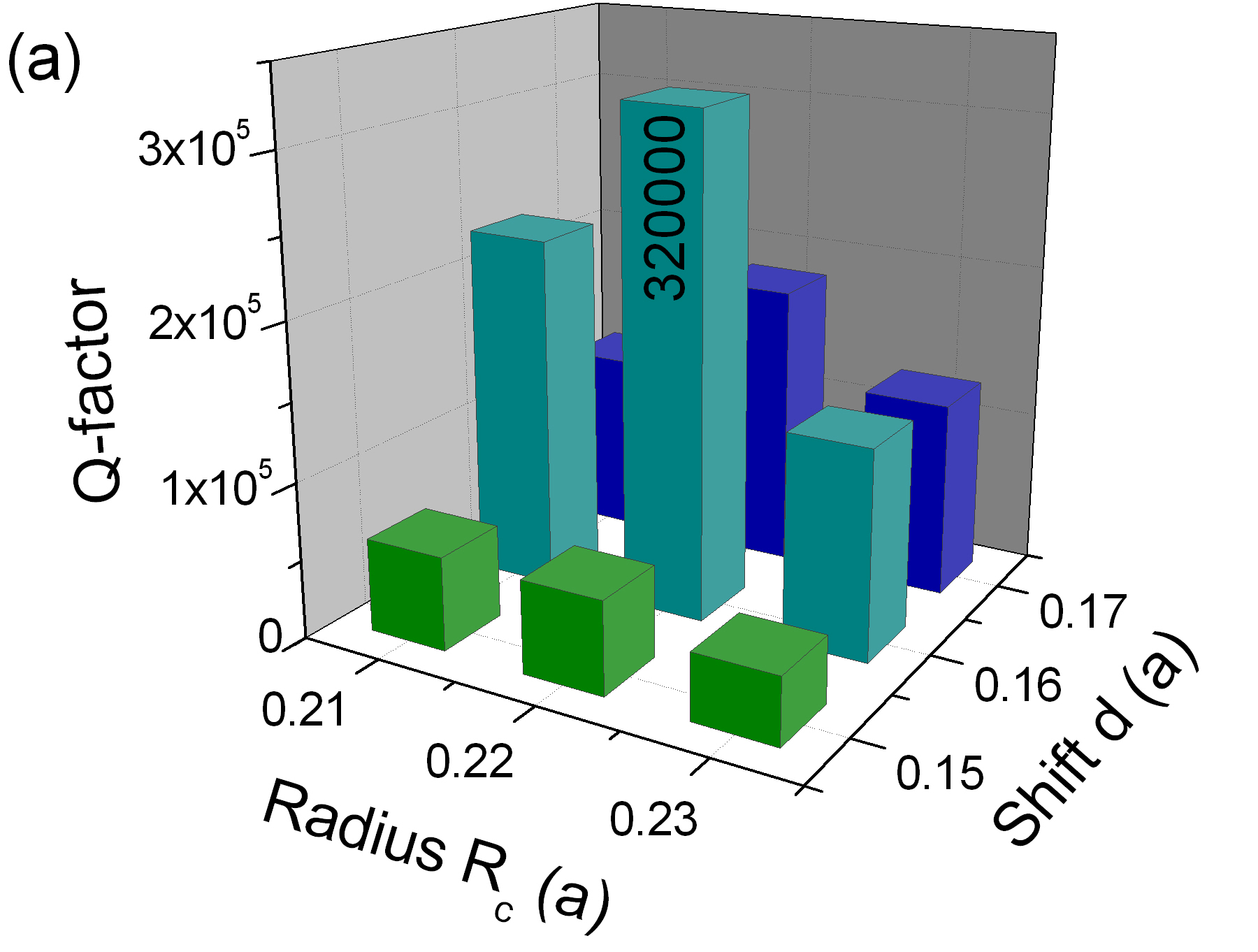}
  \label{fig:Tolerance-Q-factor}}
  \subfigure{
  \includegraphics[width=0.49\textwidth]{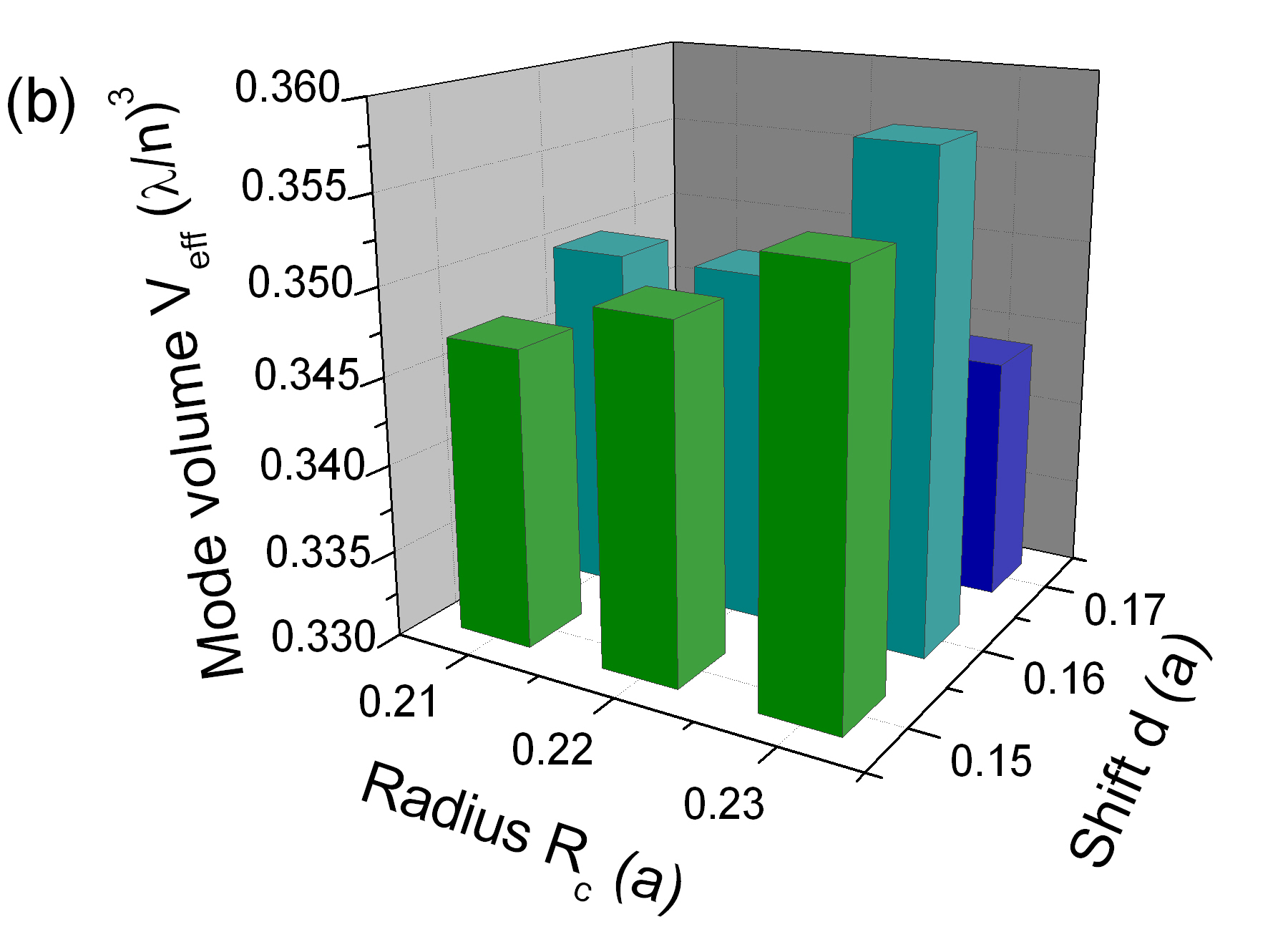}
  \label{fig:Tolerance-Modvol}}
  \caption{Fabrication tolerance test: Dependence of (a) the $Q$-factor and (b) the modal volume $V_{\text{eff}}$ of the optimized M0$^{F}$ cavity on the radius $R_c$ and the displacement $d$.\label{fig:stability}}
\end{figure}
In view of experimental realization of photonic crystals in diamond, it is interesting to investigate the influence of the cavity $Q$ on fabrication tolerances. As discussed in section \ref{sec:optimization}, the quality factor is significantly enhanced up to $Q=320000 $ by modification of the holes surrounding the defect. In this section, we investigate how sensitive the high-$Q$ modes are to variations of the neighboring holes. Exemplarily, the next-neighboring holes $b$ and $c$ are considered. Starting from the optimized M0$^F$-cavity (see Tab. \ref{tab:M0g}), we vary the displacement $d \in [0.15a, \, 0.17a]$  and the radius  $R_c \in [0.21a, \, 0.23a]$  in steps of 0.01$a$. The dependence of the quality factor and the modal volume on the surrounding structural parameters are shown in Fig. \ref{fig:stability}: The displacement $d$ seems to be a crucial parameter. If $d$ changes from $0.16a$ to $0.15a$, the $Q $ factor drops down to $4-6\cdot 10^4$, for all radii $R_c$. In contrast, if we keep the shift $d=0.16a$ of the holes $b$ fixed and vary the radius $R_c \in [0.21a, \, 0.23a]$ only, the $Q$-factor of the M0$^F$ cavity always exceeds $ 10^5$. This is favorable considering that it is  easier to fix the central position of the air holes
than to determine their radius using common etching techniques.
If we envisage a central wavelength $\lambda = $ \unit[637]{nm} of the M0$^F$-cavity equal to the emission wavelength of the NV-center, the lattice constant is $a= $ \unit[234]{nm}. To achieve large quality factors in practice, high precision  structuring techniques with fabrication tolerances smaller than \unit[3]{nm} are required.
One available technique  is to use a  focused beam of gallium ions (energy \unit[30]{keV}) to pattern the diamond membrane.
The spot size of the gallium ion beam  is in the order of several nanometers, so structures can be fabricated with a resolution of \unit[6]{nm} \cite{Daniel2001}.
Another technique which has already been used to fabricate photonic crystals in diamond \cite{Wang2007} is  reactive ion etching achieving radii of \unit[80]{nm} $\pm$ \unit[5]{nm} \cite{Wang2008a}. The radii of the milled air holes are sensitive to the dose  used in the e-beam lithography to pattern the etch mask and to the applied etching parameters.

In contrast to the cavity $Q$, the modal volume seems to be nearly independent of the fine-tuning of the surrounding holes. As shown in Fig. \ref{fig:Tolerance-Modvol}, the modal volume  varies between $V_{\text{eff}} = (0.336-0.358) (\lambda/n)^3$. This is a huge advantage of the M0 cavity design in view of achieving large enhancement of the spontaneous emission. Even if the quality factor is reduced to $Q \approx 4\cdot 10^4$ due to fabrication tolerances,  theoretical Purcell factors of $F \approx 8500$ can be obtained by coupling an emitter to the M0$^F$-cavity.

\subsection{M0 cavity design in a silicon slab}
\label{sec:Silicon}
As discussed above, we have optimized the M0-cavity structure in diamond such that vertical radiation losses have been reduced significantly yielding  a $Q$-factor even larger than predicted for M0-cavity structures in silicon \cite{Zhang2004}. Because of the larger refractive index of silicon, this suggest that further improvement of the  M0-cavity geometry can also be obtained  in a silicon slab  using the same design process as described above.
As a starting point for the optimization in silicon we choose the M0-cavity structure presented in Ref. \cite{Zhang2004} yielding a quality factor of $Q \approx 135000$ with a modal volume of $V_{\text{eff}}=0.288 (\lambda/n)^3$.
After carefully fine-tuning the background radius $R$ and the thickness $h$ of the slab  as well as the surrounding holes according to the procedure described in section \ref{sec:optimization}, we find the optimal M0$^{Si}$-structure listed in Tab. \ref{tab:M0j}. The other radii not listed in Tab. \ref{tab:M0j} correspond to the background radius $R=0.26a$. As a result, we gain an improvement in cavity $Q$ by more then a factor of three with $Q\approx 458000$ and $V_{\text{eff}}=0.260 (\lambda/n)^3$.
This result confirms that adaption of air holes even far away from the defect generally improves the cavity lifetime significantly.

\begin{table}[h]
\centering
\begin{tabular}{|c|c|c|c|c|c|c|c|c|c|c|}
\hline
{$R$}[$a$] & $h$[$a$] & {$d$}[$a$]  & {$R_c$}[$a$] &
 {$ R_k$}[$a$]  &{$R_m$}[$a$]& {$R_s$}[$a$] & {$R_t$}[$a$] &{$R_v$}[$a$]  & {$Q$}  \\
\hline
      0.26 &    0.8&   0.14 & 0.24 & 0.24  &0.25 & 0.24  & 0.28 & 0.25&  458000 \\
\hline
\end{tabular}
\caption{Optimization of the M0-cavity in a silicon slab (Parameter Set  $M0^{Si}$). The radii of the air holes not listed here are equal to the background radius $R$.
\label{tab:M0j}}
\end{table}

\section{Radiation $Q$-factors versus FDTD results}\label{sec:radiatedPower}
\begin{figure}
  \centering
  \includegraphics[width=0.5\textwidth]{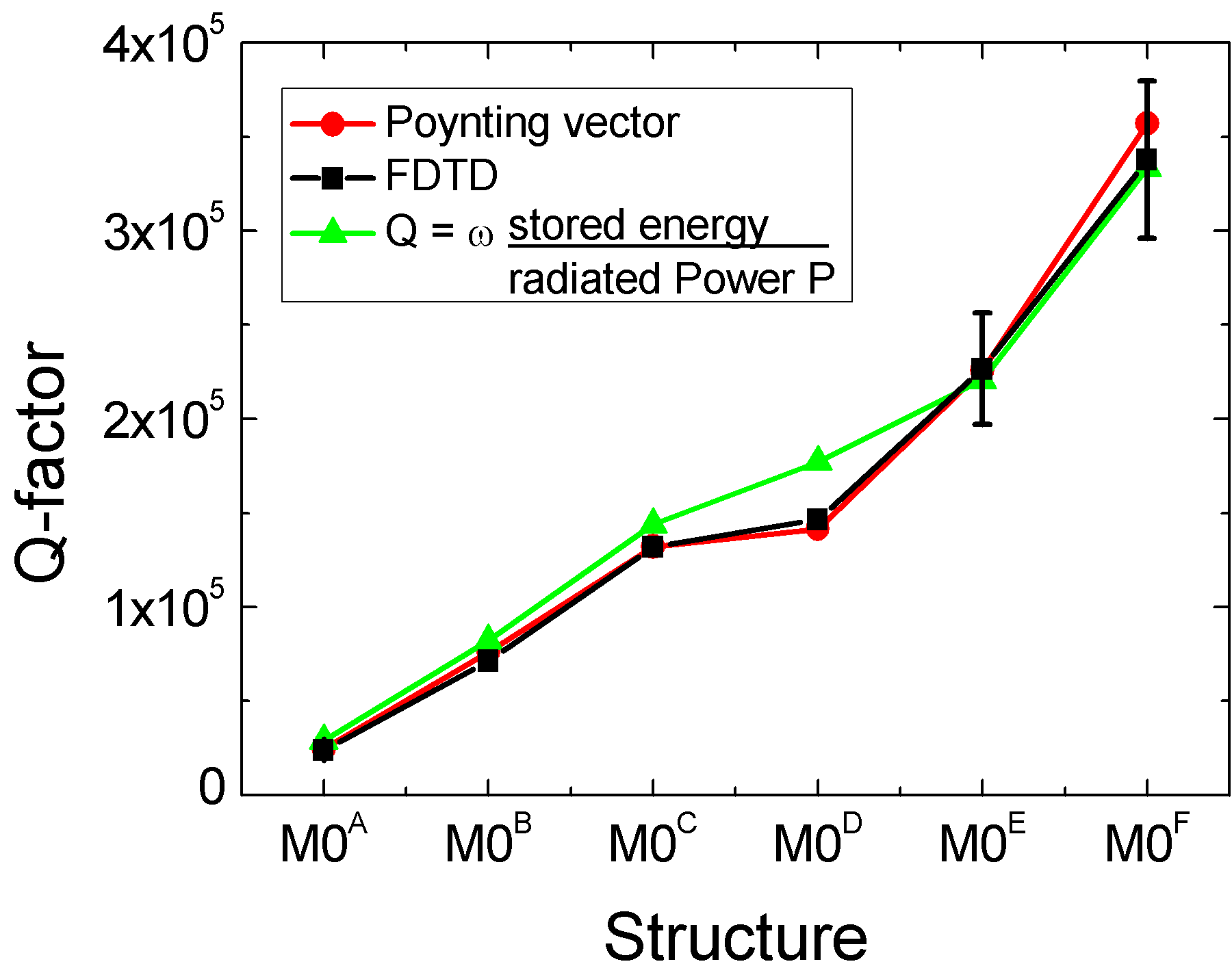}
  \caption{Comparison between the $Q$-factors calculated from the radiated power according to equation \eqref{Q_U/P} and the FDTD simulations. On the one hand the radiated power is determined by the Poynting vector (red points) and on the other hand by integrating over all components inside the light cone using equation \eqref{Fourier-vuck} (green triangles). \label{fig:Fourier-Poyn-FDTD}}
\end{figure}
In section \ref{sec:optimization}, we calculate the radiated intensity  from the 2D spatial Fourier transforms of the cavity near field. The leaky components inside the light cone indicate in which direction further improvement of the cavity geometry is necessary. Furthermore, the radiated intensity $I$ gives not only qualitative information about the radiation losses but also permits to draw quantitative conclusions. By summing over all wave vector components inside the light cone according to equation \eqref{Fourier-vuck}, we can calculate the radiated power $P$. This offers in alternative method to determine the cavity $Q$-factor according to equation \eqref{Q_U/P}: dividing the stored energy $U$ in the cavity by the vertical radiated power $P$ and multiplying by the resonance frequency $\omega$.
In this section, we compare the  radiation $Q$-factors with the FDTD results.
The $Q$-factors calculated by the FDTD algorithm might depend on the chosen resolution (black rectangles in Fig. \ref{fig:Fourier-Poyn-FDTD}). The errorbars in Fig. \ref{fig:Fourier-Poyn-FDTD} (exemplarily for structures M0$^A$, M0$^E$, M0$^F$) show the variation of the FDTD results with the change of resolution from 32 to  50 points per lattice constant. Moreover, the application of mirror symmetries to reduce the computational cell, might also cause an overestimation of the $Q$-factors \cite{Vuckovic2001}.
Therefore, it is interesting to have a simple alternative method to determine the $Q$-factors in order to identify numerical artifacts.

We calculate the  near field patterns as well as the stored energy $U$ in the cavity using FDTD simulations.
The vertical radiation losses are determined via the Fourier transforms of the near field (see Sec. \ref{sec:ComputationMethod}). The spatial Fourier transforms are calculated in a plane $S$ at position $ z = \lambda/2$. To determine the vertical radiated power, the integral \eqref{Fourier-vuck} over all wave vector components inside the light cone is performed as a discrete sum.
Fig. \ref{fig:Fourier-Poyn-FDTD} shows a comparison between the $Q$-factors calculated via radiated power (green triangles) and the FDTD results (black rectangles) for the cavity designs in diamond  presented in sections \ref{sec:M0Cavity} and \ref{sec:optimization}.
This simple calculation procedure reproduces well the FDTD results for different cavity designs. For structures M0$^A$, M0$^B$, M0$^C$ and M0$^D$ with small background radius $R=0.26a$, deviations from the FDTD results are maximally 20\% (structure M0$^D$). These deviations are mainly due to the fact, that the resonance frequency is close to the upper edge of the photonic band gap leading to imperfect lateral  confinement of the cavity mode. Albeit, the radiation $Q$-factor, taking into account vertical radiation losses only, reflects well the improvement of the cavity lifetime by optimizing the photonic crystal geometry.   For structures M0$^E$ and M0$^F$  with enlarged background radii $R = 0.28a$, the deviations of the radiation $Q$-factor from the FDTD results are smaller than 6\%. With increasing background radii, the photonic band gap increases and the resonance frequency of the cavity mode lies deeper inside the photonic band gap resulting in a better lateral mode confinement.
Therefore, by calculating the vertical radiation losses via Fourier transform of the near field, the radiation $Q$-factor of  photonic crystal geometries with enlarged background radii is a good approximation to the FDTD results.

A third approach for determining the $Q$-factor uses the radiated power as well as the stored energy obtained from FDTD simulations. The total radiated power in this case is computed by the integral of the Poynting vector $\mathfrak{Re}(\vec{E}^{*} \times \vec{H})$ in all three directions through planes at the edge of the simulation cell. The associated $Q$-factors for   cavity geometries in diamond  are shown by the red points in Fig. \ref{fig:Fourier-Poyn-FDTD}: The agreement with the FDTD results is very well.

Calculating the stored energy as well as the radiated power, either using FDTD simulations or via  spatial Fourier transforms of the near field, offers an alternative way to determine the cavity $Q$-factor. This method can be used to check the results of the FDTD simulations, usually obtained from the analysis of the cavity temporal decay, and to easily identify  possible numerical artifacts.

\section{Absorption}\label{sec:Absorption}

For the practical realization of photonic crystal microcavities in diamond it is essential to investigate the influence of the material absorption on the cavity $Q$-factor.
To achieve effective optical coupling in practice, ultrapure, high-quality diamond films are required. Actually, our measurements show \cite{Hepp2009}, that nano-crystalline diamond films strongly absorb light at the emission wavelengths of the NV, SiV and NE8 color center ($\sim$ \unit[630]{nm} - \unit[800]{nm})  due to graphite or amorphous carbon in the grain boundaries \cite{Achatz2006}. The absorption coefficient is determined by measuring the transmission through a 
nano-crystalline diamond membrane ($\rho$-BeSt, Innsbruck, Austria). The reduction of the transmission signal due to surface roughness  (\unit[15]{nm} rms) is estimated to  2\%, while the material absorption leads to a reduction  of the signal of 11\% (for a film of \unit[300]{nm} thickness). Therefore, we assume that material absorption of the dielectric background is the primary loss mechanism of microcavities in photonic crystals in nano-crystalline diamond films.

\begin{figure}
  \centering
  \includegraphics[width=0.45\textwidth]{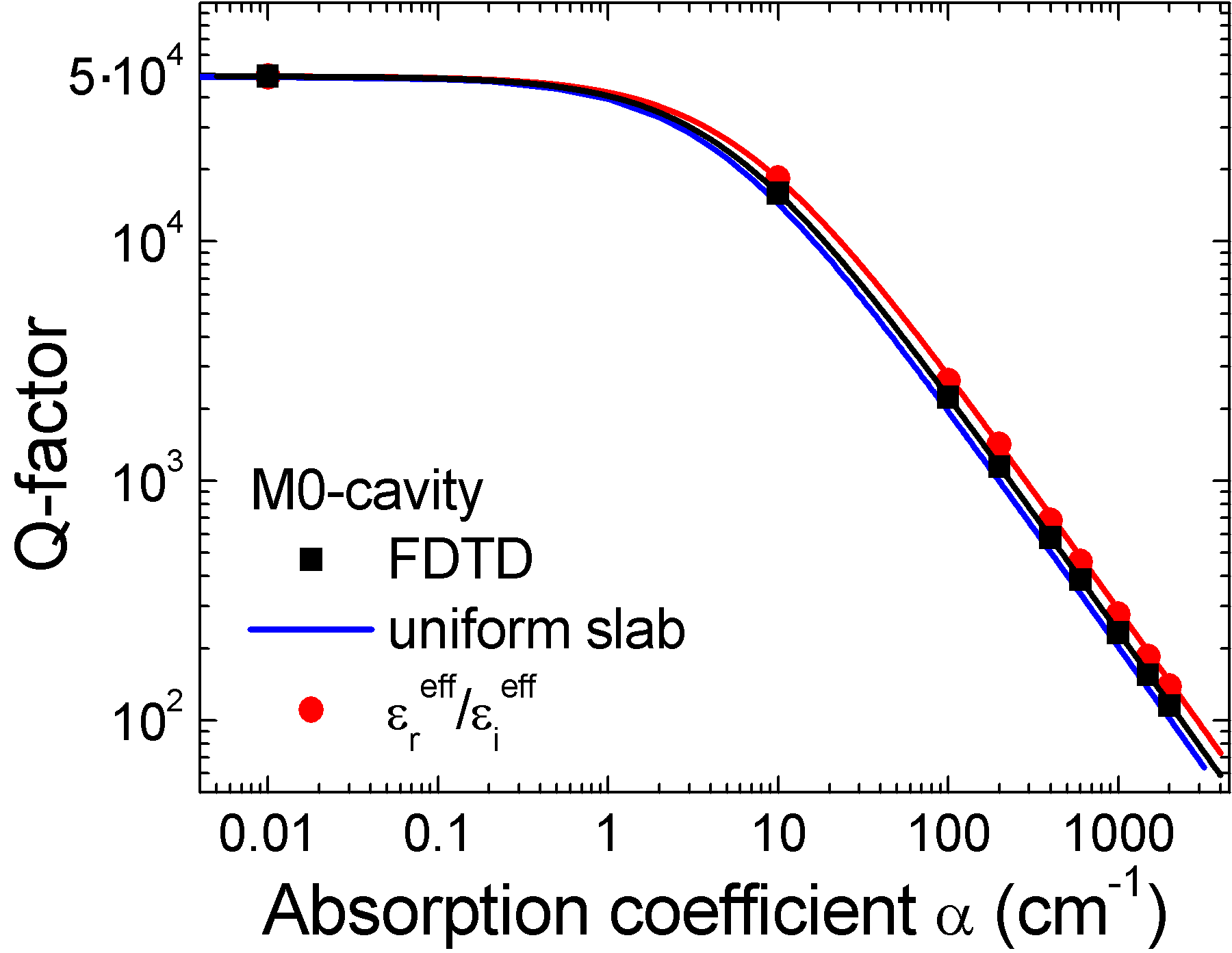}
  \caption{Influence of the absorption coefficient $\alpha$ of the diamond material on the $Q$-factor of the M0-cavity ($R= 0.26a$, $d=0.15a$, $R_c=R_e=0.24a$, $h=0.91a$): The FDTD results are shown in black. The dependance can be described by a simple model of a linear cavity filled with a homogenous diamond slab or filled with the actual photonic crystal structure.\label{fig:AbsorptionM0}}
\end{figure}

To investigate the influence of material absorption on the cavity $Q$-factor, we implement artificial resonances in the dielectric function such that the imaginary part $\epsilon_i$ yields the desired resonant absorption whereas the real part $\epsilon _r$ remains almost constant \cite{Kreuzer2008}. The results of the FDTD simulations (black rectangles in  Fig. \ref{fig:AbsorptionM0}) reveal that the $Q$-factor strongly depends on the absorption coefficient $\alpha $ of the diamond material. For our ultra-nano-crystalline diamond films we measure maximum absorption coefficients $\alpha \approx $ \unit[4000]{cm$^{-1}$} at the emission wavelength $\lambda=$ \unit[637]{nm} of the NV center \cite{Hepp2009}.
Starting from a theoretical quality factor $Q \approx 50000$ of a M0-cavity with optimized next neighboring holes ($R= 0.26a$, $d=0.15a$, $R_c=R_e=0.24a$, $h=0.91a$), the material absorption of the diamond film would reduce the quality factor to only $Q<100$.
Nevertheless,  coupling an emitter to a M0-cavity, a Purcell-Factor of about $F=15$ is still  achievable because of the extraordinary small modal volume.
Due to recent progress in growing high-quality diamond films by chemical-vapor deposition \cite{Schreck1999},   absorption coefficients of $\alpha \approx $ \unit[200]{cm$^{-1}$} \cite{Schreck2009} can be obtained. With such low loss diamond membranes an experimental cavity $Q$-factor  of 1150 can be achieved, yielding a Purcell-Factor of $F=220$.

Calculation of  the quality factors for different absorption coefficients $\alpha$ using FDTD simulations is a very time-consuming task. Therefore, an analytical approximation to  evaluate  the $Q$-factor for lossy materials would be very helpful. The dependence of  $Q$ on $\alpha$ can be described by a simple model: In the case of a linear cavity with a homogenous diamond slab  between two  mirrors the total quality factor $Q$ is given by \cite{Asano2006, Xu2007}:
%
\begin{equation}
    Q^{-1}= Q_0{^{-1}} + Q_{\text{abs}}^{-1},
    \label{Q-Absorption}
\end{equation}
where $Q_0$ denotes the quality factor of an ideal, lossless cavity and $Q_{\text{abs}}$  the quality factor due to material absorption. As described in Refs. \cite{Kreuzer2008, Xu2007}, for weakly absorbing material $\epsilon_r \gg \epsilon_i$ the absorption quality factor can be written as $Q_{\text{abs}}= k/\alpha$, where the absorption coefficient is given by $\alpha= k_0 \epsilon_i /\sqrt{\epsilon_r}$. The wave vector in the dielectric is $k= \sqrt{\epsilon_r} k_0$, where $k_0$ denotes the wave vector in vacuum.
%
Under the assumption that most of the electric field is concentrated in the dielectric, the material losses are given by:
\begin{equation}
    Q_{\text{abs}}= k/\alpha = \epsilon_r/\epsilon_i
    \label{Qabs-homoSchicht}
\end{equation}
The blue curve in Fig. \ref{fig:AbsorptionM0} shows that this first approximation predicts well the dependance of the cavity $Q$-factor. The deviations from the FDTD results are smaller than 13\%. Actually, equation  \eqref{Qabs-homoSchicht} denotes a lower limit to the FDTD results for lossy materials, because the electric field is not completely concentrated within the dielectric but also leaks into the air holes.

In a more general approach, we want to take into account the actual photonic crystal structure. For weakly absorbing material $\epsilon_r \gg \epsilon_i$ the  imaginary part $\epsilon_i$ can be considered  a small perturbation $\Delta \epsilon = i \epsilon_i $ of the dielectric function $\epsilon(\vec{r})$. This perturbation $\Delta \epsilon$ results in a first-order correction of the  resonance frequency $\Delta \omega= i \omega_i $ \cite{Johnson2002, Joannopoulos2008}:
%
\begin{equation}
    \Delta \omega = -\frac{\omega_r}{2}
    \frac{\int \Delta \epsilon (\vec{r})\,|\vec{E}(\vec{r})|^2 d^3r}{\int \epsilon_r(\vec{r})\,|\vec{E}(\vec{r})|^2 d^3r},
    \label{imaginärTeil-omega}
\end{equation}
where $\omega_r$ denotes the resonance frequency of the ideal, lossless cavity structure. Considering that the $Q$-factor is defined as $Q= -\frac{\omega_r}{2\omega_i}$, the absorption quality factor can be written as:
%
\begin{equation}
    Q'_{\text{abs}} = \frac{\int \epsilon_r (\vec{r})\,|\vec{E}(\vec{r})|^2 d^3r}{\int \epsilon_i(\vec{r})\,|\vec{E}(\vec{r})|^2 d^3r}
    \label{Qabs-Störungstheorie}
\end{equation}
The integral in equation \eqref{Qabs-Störungstheorie} is taken over the entire photonic crystal slab with the  electric field $\vec{E}(\vec{r})$ of the ideal, lossless photonic crystal structure as a weighting factor. Therefore the electric field has to be calculated only once for  an ideal cavity  using numerical methods such as FDTD simulations. Afterwards it can be reused to calculate $Q'_{\text{abs}}$  for different lossy materials.   The overlap integral in equation \eqref{Qabs-Störungstheorie} of the electric field and the dielectric function is performed as a discrete sum.  The red curve in Fig. \ref{fig:AbsorptionM0} shows the absorption quality factor $Q'_{\text{abs}}$ calculated for different absorption coefficients $\alpha$ of the dielectric background.
 The deviations from the FDTD results are mostly due to  different procedures used to calculate the overlap integral in equation \eqref{Qabs-Störungstheorie}. The discrete summation leads to a slightly higher $Q$-factor of about 11\% compared to an exact integration. We estimate the error by calculating the effective dielectric constant $\int \epsilon(\vec{r}) |\vec{E}(\vec{r})|^2 d^3r / \int |\vec{E}(\vec{r})|^2 d^3r $ performing the integrals as a discrete sum and comparing it to FDTD simulations.

Both simple models obtained above predict well the dependence of the quality factor for lossy materials. These simple equations permit to calculate an upper and a lower limit to the FDTD results without relying on time-consuming simulations.

\section{Conclusion}

We have presented an optimized M0-microcavity design in a diamond-based photonic crystal. The small defect is introduced by shifting two holes outwards along the  $x$-axis yielding  an extremely small modal volume of $V_{\text{eff}}=0.35 (\lambda/n)^3$. Using Fourier- and real-space analysis, we improved the cavity structure by systematically varying the radii of the neighboring holes around the defect. The improved design yields a quality factor of $Q =320000$. Coupling a color center to such an optimized cavity structure leads to a theoretical Purcell-Factor of $F=69500$, which is one of the largest enhancement of the spontaneous emission that has been predicted so far for photonic crystal point defects in diamond. Using the same design process, we have additionally optimized the M0-cavity in silicon gaining a factor of three in cavity $Q$. The significant improvement of the cavity lifetime by  ``gentle confinement''  shows, that it is worth looking closer at the surrounding holes even far away from the defect.

In view of the practical realization of photonic crystals in nano-crystalline diamond films, material absorption of the dielectric background has to be taken into account.
The influence of absorption losses on the cavity $Q$-factor can be described by a simple model of a linear cavity: In first approximation, we consider a homogenous diamond slab between two perfect mirrors. This simple description  denotes a lower limit to the FDTD results, because the electric field is not completely concentrated within the dielectric, but also leaks into the air holes. The second approximation takes into account the actual photonic crystal structure. These two simple models offer an efficient way to predict quality factors of photonic crystal cavities for lossy materials without relying on time-consuming  FDTD simulations. Whether this simple model of a linear cavity is valid for a larger range of defect geometries in photonic crystals is subject of current work. Preliminary results show that  equation \eqref{Qabs-homoSchicht} predicts the absorption quality factor $Q_{\text{abs}}$ even better for large photonic crystal point defects.

One possibility to overcome the limitation of the $Q$-factor by material absorption is the use of single crystal diamond films. Recent progress in fabrication of thin free-standing single crystal diamond membranes, might pave the way for efficient direct coupling of color centers to high-$Q$ diamond-based photonic crystal microcavities.

\section{Acknowledgements}
The authors thank C. Hepp for the absorption measurements of our diamond films and D. Steinmetz for valuable help with the computing infrastructure. E. Neu acknowledges support from the Stiftung der Deutschen Wirtschaft (SDW). This work is funded by the Deutsche Forschungsgemeinschaft (DFG).


\begin{thebibliography}{10}
\expandafter\ifx\csname url\endcsname\relax
  \def\url#1{\texttt{#1}}\fi
\expandafter\ifx\csname urlprefix\endcsname\relax\def\urlprefix{URL }\fi
\expandafter\ifx\csname href\endcsname\relax
  \def\href#1#2{#2} \def\path#1{#1}\fi

\bibitem{Englund2005}
D.~Englund, D.~Fattal, E.~Waks, G.~Solomon, B.~Zhang, T.~Nakaoka, Y.~Arakawa,
  Y.~Yamamoto, J.~Vu\v{c}kovi\'{c}, Controlling the spontaneous emission rate
  of single quantum dots in a two-dimensional photonic crystal, Phys. Rev.
  Lett. 95 (2005) 013904.

\bibitem{Strauf2006}
S.~Strauf, K.~Hennessy, M.~T. Rakher, Y.-S. Choi, A.~Badolato, L.~C. Andreani,
  E.~L. Hu, P.~M. Petroff, D.~Bouwmeester, Self-tuned quantum dot gain in
  photonic crystal lasers, Physical Review Letters 96 (2006) 127404.

\bibitem{Noda2006}
S.~Noda, Seeking the ultimate nanolaser, Science 314 (2006) 260--261.

\bibitem{Nomura2006}
M.~Nomura, S.~Iwamoto, K.~Watanabe, N.~Kumagai, Y.~Nakata, S.~Ishida,
  Y.~Arakawa, Room temperature continuous-wave lasing in photonic crystal
  nanocavity, Opt. Express 14 (2006) 6308--6315.

\bibitem{Takano2006}
H.~Takano, B.-S. Song, T.~Asano, S.~Noda, Highly efficient multi-channel drop
  filter in a two-dimensional hetero photonic crystal, Opt. Express 14 (2006)
  3491--3496.

\bibitem{Yoshie2004}
T.~Yoshie, A.~Scherer, J.~Hendrickson, G.~Khitrova, H.~M. Gibbs, G.~Rupper,
  O.~B. Shchekin, D.~G. Deppe, Vacuum rabi splitting with a single quantum dot
  in a photonic crystal nanocavity, Nature 432 (2004) 200.

\bibitem{Hennessy2007}
K.~Hennessy, A.~Badolato, M.~Winger, D.~Gerace, M.~Atat\"{u}re, S.~Gulde,
  S.~F\"{a}lt, E.~L. Hu, A.~Imamo\v{g}lu, Quantum nature of a strongly coupled
  single quantum dot-cavity system, Nature 445 (2007) 896.

\bibitem{Khitrova2006}
G.~Khitrova, H.~M. Gibbs, M.~Kira, S.~W. Koch, A.~Scherer, Vacuum rabi
  splitting in semiconductors, Nature Phys. 2 (2006) 81--90.

\bibitem{Prawer2008}
S.~Prawer, A.~D. Greentree, Diamond for quantum computing, Science 320 (2008)
  1601.

\bibitem{Greentree2008}
A.~D. Greentree, B.~A. Fairchild, F.~M. Hossain, S.~Prawer, Diamond integrated
  quantum photonics, Materials Today 11 (2008) 22--31.

\bibitem{Su2008}
C.-H. Su, A.~D. Greentree, L.~C.~L. Hollenberg, Towards a picosecond
  transform-limited nitrogen-vacancy based single photon source, Opt. Express
  16 (2008) 6240--6250.

\bibitem{Young2009}
A.~Young, C.~Y. Hu, L.~Marseglia, J.~P. Harrison, J.~L. O'Brien, J.~G. Rarity,
  Cavity enhanced spin measurement of the ground state spin of an {NV} center in
  diamond, New Journal of Physics 11 (2009) 013007.

\bibitem{Childress2006}
L.~Childress, J.~M. Taylor, A.~S. S{\o}rensen, M.~D. Lukin, Fault-tolerant
  quantum communication based on solid-state photon emitters, Phys. Rev. Lett.
  96 (2006) 070504.

\bibitem{Greentree2006}
A.~Greentree, J.~Salzman, S.~Prawer, L.~Hollenberg, Quantum gate for \textsc{Q}
  switching in monolithic photonic-band-gap cavities containing two-level
  atoms, Phys. Rev. A 73 (2006) 013818.

\bibitem{lim:012304}
Y.~L. Lim, S.~D. Barrett, A.~Beige, P.~Kok, L.~C. Kwek, Repeat-until-success
  quantum computing using stationary and flying qubits, Phys. Rev. A 73 (2006)
  012304.

\bibitem{Benjamin2009}
S.~C. Benjamin, B.~W. Lovett, J.~M. Smith, Prospects for measurement-based
  quantum computing with solid state spins, Laser \& Photon. Rev.

\bibitem{Purcell1946}
E.~Purcell, Spontaneous emission probabilities at radio frequencies, Phys. Rev.
  Lett. 69 (1946) 681.

\bibitem{Kreuzer2008}
C.~Kreuzer, J.~Riedrich-M\"{o}ller, E.~Neu, C.~Becher, Design of photonic
  crystal microcavities in diamond films, Opt. Express 16 (2008) 1632--1644.

\bibitem{Bong-Shik2005}
B.-S. Song, S.~Noda, T.~Asano, Y.~Akahane, Ultra-high-\textsc{Q} photonic
  double-heterostructure nanocavity, Nature 4 (2005) 207 -- 210.

\bibitem{Eiichi2006}
E.~Kuramochi, M.~Notomi, S.~Mitsugi, A.~Shinya, T.~Tanabe, Ultrahigh-\textsc{Q}
  photonic crystal nanocavities realized by the local width modulation of a
  line defect, Appl. Phys. Lett. 88 (2006) 041112.

\bibitem{Takahashi2007}
Y.~Takahashi, H.~Hagino, Y.~Tanaka, B.-S. Song, T.~Asano, S.~Noda, High-\textsc{Q}
  nanocavity with a 2-ns photon lifetime, Opt. Express 15 (2007) 17206.

\bibitem{Bayn2006}
I.~Bayn, J.~Salzman, High-\textsc{Q} photonic crystal nanocavities on diamond
  for quantum electrodynamics, Eur. Phys. J. Appl. Phys. 37 (2007) 19--24.

\bibitem{Bayn2008}
I.~Bayn, J.~Salzman, Ultra-high-\textsc{Q} photonic crystal nanocavity design: The
  effect of a low-$\epsilon$ slab material, Opt. Express 16 (2008) 4972--4980.

\bibitem{Tomljenovic-Hanic2009}
S.~Tomljenovic-Hanic, A.~D. Greentree, C.~M. de~Sterke, S.~Prawer, Flexible
  design of ultrahigh-\textsc{Q} microcavities in diamond-based photonic crystal slabs,
  Opt. Express 17 (2009) 6465--6475.

\bibitem{Wang2007}
C.~F. Wang, R.~Hanson, E.~Hu, T.~Feygelson, J.~Yang, J.~E. Butler, Fabrication
  and characterization of two-dimensional photonic crystal microcavities in
  nanocrystalline diamond, Appl. Phys. Lett. 91 (2007) 201112.

\bibitem{Zhang2004}
Z.~Zhang, M.~Qiu, Small-volume waveguide-section high \textsc{Q} microcavities
  in \textsc{2D} photonic crystal slabs, Opt. Express 12 (2004) 3988--3995.

\bibitem{Yoshihiro2003}
Y.~Akahane, T.~Asano, B.-S. Song, S.~Noda, High-\textsc{Q} photonic nanocavity
  in a two-dimensional photonic crystal, Nature 425 (2003) 944 -- 947.

\bibitem{Yoshihiro2005}
Y.~Akahane, T.~Asano, B.-S. Song, S.~Noda, Fine tuned high-\textsc{Q} photonic
  crystal nanocavity, Opt. Express 13 (2005) 1202 -- 1214.

\bibitem{S.2006}
S.~Tomljenovic-Hanic, M.~J. Steel, C.~M. Sterke, J.~Salzman, Diamond based
  photonic crystal microcavities, Opt. Express 14 (2006) 3556 -- 3562.

\bibitem{Olivero2005}
P.~Olivero, S.~Rubanov, P.~Reichart, B.~Gibson, S.~Huntington, J.~Rabeau,
  A.~Greentree, J.~Salzman, D.~Moore, D.~Jamieson, S.~Prawer, Ion-beam-assisted
  lift-off technique for three-dimensional micromachining of freestanding
  single-crystal diamond, Adv. Mat. 17 (2005) 2427--2430.

\bibitem{Olivero2006}
P.~Olivero, S.~Rubanov, P.~Reichart, B.~Gibson, S.~Huntington, A.~Rabeau,
  A.~Greentree, J.~Salzman, D.~Moore, D.~Jamieson, S.~Prawer, Characterization
  of three-dimensional microstructures in single-crystal diamond, Diamond
  Relat. Mater. 15 (2006) 1614--1621.

\bibitem{Fairchild2008}
B.~A. Fairchild, P.~Olivero, S.~Rubanov, A.~Greentree, F.~Waldermann, R.~A.
  Taylor, J.~M. Smith, S.~Huntington, B.~Gibson, D.~Jamieson, S.~Prawer,
  Fabrication of ultrathin single-crystal diamond membranes, Adv. Mater. 20
  (2008) 1--6.

\bibitem{Hiscocks2008}
M.~P. Hiscocks, C.~J. Kaalund, F.~Ladouceur, S.~Huntington, B.~C. Gibson,
  S.~Trpkovski, D.~Simpson, E.~Ampem-Lassen, S.~Prawer, J.~E. Butler, Reactive
  ion etching of waveguide structures in diamond, Diamond Relat. Mater. 17
  (2008) 1831--1834.

\bibitem{Hiscocks2008a}
M.~P. Hiscocks, K.~Ganesan, B.~C. Gibson, S.~T. Huntington, F.~Ladouceur,
  S.~Prawer, Diamond waveguides fabricated by reactive ion etching, Opt.
  Express 16 (2008) 19512--19519.

\bibitem{Taflove1998}
A.~Taflove, Advances in Computational Electrodynamics, Artech House Inc.,
  Norwood, 1998.

\bibitem{Farjadpour2006}
A.~Farjadpour, D.~Roundy, A.~Rodriguez, M.~Ibanescu, P.~Bermel, J.~D.
  Joannopoulos, S.~G. Johnson, G.~Burr, Improving accuracy by subpixel
  smoothing in \textsc{FDTD}, Opt. Lett. 31 (2006) 2972 -- 2974.

\bibitem{Berenger1994}
J.-P. Berenger, A perfectly matched layer for the absorption of electromagnetic
  waves, J. Comp. Phys. 114 (1994) 185--200.

\bibitem{Mandelshtam1997}
V.~A. Mandelshtam, H.~S. Taylor, Harmonic inversion of time signals and its
  applications, J. Chem. Phys. 107 (1997) 6756--6769.

\bibitem{Vuckovic2001}
J.~Vu\v{c}kovi\'{c}, M.~Lon\v{c}ar, H.~Mabuchi, A.~Scherer, Design of photonic
  crystal microcavities for cavity \textsc{QED}, Phys. Rev. E 65 (2001) 016608.

\bibitem{Dirk2005}
D.~Englund, I.~Fushman, J.~Vu\v{c}kovi\'{c}, General recipe for designing
  photonic crystal cavities, Opt. Express 13 (2005) 5961 -- 5975.

\bibitem{Vuckovic2002b}
J.~Vu\v{c}kovi\'{c}, M.~Lon\v{c}ar, H.~Mabuchi, A.~Scherrer, Optimization of
  the \textsc{Q} factor in photonic crystal microcavities, IEEE J. Quantum.
  Electron. 38 (2002) 850 -- 856.

\bibitem{Zaitsev2001}
A.~M. Zaitsev, Optical Properties of Diamond, Springer-Verlag, Berlin, 2001.

\bibitem{Daniel2001}
G.~Daniel, C.~Ulrich, R.~Warschauer, Documentation xP DualBeam$^{\text{TM}}$
  Workstation Strata, FEI Company (2001).

\bibitem{Wang2008a}
C.-F. Wang, Design, fabrication and characterization of diamond-based photonic
  microcavities, Ph.D. thesis, University of California, Santa Barbara (March
  2008).

\bibitem{Hepp2009}
C.~Hepp, D.~Steinmetz, E.~Neu, C.~Becher, S.~Ghodbane, D.~Steinm\"{u}ller,
  Optical properties of nanocrystalline diamond films for photonic crystals
  microcavities, in preparation (2009).

\bibitem{Achatz2006}
P.~Achatz, J.~Garrido, M.~Stutzmann, O.~Williams, Optical properties of
  nanocrystalline diamond thin films, Appl. Phys. Lett. 88 (2006) 101908.

\bibitem{Schreck1999}
M.~Schreck, H.~Roll, B.~Stritzker, Diamond/{I}r/{S}r{T}i{O}$_3$: A material combination
  for improved heteroepitaxial diamond films, App. Phys. Lett. 74 (1999)
  650--652.

\bibitem{Schreck2009}
M.~Schreck, {L}ehrstuhl f\"ur Experimentalphysik IV, Universit\"at Augsburg,
  Universit\"atsstr. 1, 86135 Augsburg (personal communication, 2009).

\bibitem{Asano2006}
T.~Asano, B.-S. Song, S.~Noda, Analysis of the experimental \textsc{Q} factors ($\sim$ 1
  million) of photonic crystal nanocavities, Opt. Express 14 (2006) 1996--2002.

\bibitem{Xu2007}
T.~Xu, S.~Yang, S.~V. Nair, H.~E. Ruda, Nanowire-array-based photonic crystal
  cavity by finite-difference time-domain calculations, Phys. Rev. B 75 (2007)
  125104.

\bibitem{Johnson2002}
S.~G. Johnson, M.~Ibanescu, M.~A. Skorobogatiy, O.~Weisberg, J.~D.
  Joannopoulos, Y.~Fink, Perturbation theory for maxwell's equations with
  shifting material boundaries, Phys. Rev. B 65 (2002) 066611.

\bibitem{Joannopoulos2008}
J.~D. Joannopoulos, S.~G. Johnson, R.~D. Meade, J.~N. Winn, Photonic Crystals -
  Molding the Flow of Light, 2nd Edition, Princeton University Press,
  Princeton, 2008.
\end{thebibliography}
\end{document}